\documentclass[11pt,a4paper,reqno]{amsart} 

\usepackage{amssymb,graphicx}

\usepackage{amssymb}

\newcommand{\koniec}{\begin{flushright}  $\Box $ \end{flushright}}
\newtheorem{theo}{Theorem}[section]

\def\theequation{\thesection.\arabic{equation}}

\newcounter{mnotecount}[section]

\renewcommand{\themnotecount}{\thesection.\arabic{mnotecount}}

\newcommand{\mnote}[1]
{\protect{\stepcounter{mnotecount}}$^{\mbox{\footnotesize
$
\bullet$\themnotecount}}$ \marginpar{
\raggedright\tiny\em
$\!\!\!\!\!\!\,\bullet$\themnotecount: #1} }

\newcommand{\hook}{{\setlength{\unitlength}{11pt}   
                   \begin{picture}(.833,.8)
                   \put(.15,.08){\line(1,0){.35}}
                   \put(.5,.08){\line(0,1){.5}}
                   \end{picture}}}
\newcommand{\CP}{\mathbb{CP}}
\newcommand{\C}{\mathbb{C}}
\newcommand{\Z}{\mathbb{Z}}
\newcommand{\PP}{\mathbb{P}}

\newcommand{\R}{\mathbb{R}}

\def\p{\partial}
\def\OO{\mathcal{O}}

\def\ov{\overline}
\def\ve{\varepsilon}
\def\be{\begin{equation}}
\def\ee{\end{equation}}

\def\bea{\begin{eqnarray}}
\def\eea{\end{eqnarray}}

\newcommand{\spp}{\mathbb{S}}

\def\ov{\overline}

\begin{document}\date{13 June  2015}
\title{Non-relativistic twistor theory and Newton--Cartan geometry}
\author{Maciej Dunajski}
\address{Department of Applied Mathematics and Theoretical Physics\\ 
University of Cambridge\\ Wilberforce Road, Cambridge CB3 0WA\\ UK.}
\email{m.dunajski@damtp.cam.ac.uk}
\author{James Gundry}
\address{Department of Applied Mathematics and Theoretical Physics\\ 
University of Cambridge\\ Wilberforce Road, Cambridge CB3 0WA\\ UK.}
\email{J.M.Gundry@damtp.cam.ac.uk}
\begin{abstract} 
We develop a non--relativistic twistor theory, in which 
Newton--Cartan structures
of Newtonian gravity correspond to  complex three--manifolds with a four--parameter family of rational curves with normal bundle 
${\mathcal O}\oplus{\mathcal O}(2)$. 
We show that the Newton--Cartan space-times are unstable under the general
 Kodaira deformation of the twistor complex structure. The  
Newton--Cartan connections can nevertheless be reconstructed 
from Merkulov's generalisation of the Kodaira map augmented by a choice of a
holomorphic line bundle over the twistor space 
trivial on twistor lines. The Coriolis force may be incorporated by
holomorphic vector bundles, which in general are non--trivial on twistor lines. 
The resulting geometries agree with non--relativistic limits of 
anti-self-dual gravitational instantons.
\end{abstract}   
\maketitle
\section{Introduction}
Over the last six years there has been large interest
in the AdS/CFT correspondences providing the gravity
duals of non--relativistic gauge theories
relevant in  solid--state physics \cite{hartnoll}.
A non-relativistic field theory should have  a dual made out of a non-relativistic theory of gravity.
The mathematical structure of such a theory - the Newton--Cartan (NC) space-time -  is rather baroque.
It consists of a torsion--free non--metric connection $\nabla$, a parallel
degenerate metric $h$ and a parallel 
one--form $\theta$ in the kernel of this metric \cite{cartan,Trautman,duval2}.

In this paper we shall construct a 
non-relativistic twistor theory for Newton-Cartan space-times.
There are several equivalent definitions of twistors for flat Minkowski 
space, but they lead to nonequivalent pictures once a non-relativistic limit 
is taken. One could define twistors as spinors for the conformal group
$SU(2, 2)$ and then take a Wigner contraction corresponding to a limit where
the speed of light becomes infinite \cite{Luk}. This approach has to be 
abandoned in the search for an analogue of a curved
twistor correspondence \cite{penrose} as a general space time does not admit any conformal isometries. Thus the linear/group theoretic structures of the twistor space become irrelevant. We shall instead concentrate on the holomorphic geometry of the
twistor space and its one-dimensional complex submanifolds. 
In the non-relativistic limit the normal bundle of the twistor curves
corresponding to space-time points jumps from
${\mathcal O}(1)\oplus{\mathcal O}(1)$ to 
${\mathcal O}\oplus{\mathcal O}(2)$,
where ${\mathcal O}(n)$ is the holomorphic line bundle over $\CP^1$ with
the Chern class $n$. This results in the conformal structure becoming degenerate.
We shall show that the Newtonian space-times are unstable under the general
 Kodaira deformation of the twistor complex structure. 
 
 The  
Newton--Cartan connections can nevertheless be reconstructed 
from Merkulov's generalisation of the Kodaira map augmented by a choice of a
holomorphic line bundle over the twistor space  
trivial on twistor lines. The Coriolis force may be incorporated by
holomorphic vector bundles, which in general are non--trivial on twistor lines. 
The resulting geometries agree with non--relativistic limits of 
anti-self-dual gravitational instantons.
 
 The paper is organised as follows: In the next Section we shall review the Newton-Cartan
geometry, and show that in the presence of the time-independent Coriolis force, the general
Newton--Cartan connection satisfying the Einstein equations is determined by
a couple of harmonic functions on $\R^3$. In Section \ref{nr_limit}
we shall explain how the NC structures arise as limits from general relativity.
We shall give examples of NC limit of the Lorentzian 
Taub-Nut metric, anti--self--dual
(ASD) gravitational waves, and Gibbons--Hawking gravitational instantons. The one--parameter family of anti--self--dual Ricci flat metrics  
\be
\label{GH_intro}
g_{\epsilon}=(1+\epsilon V)(dx^2+dy^2+dz^2)+\frac{1}{\epsilon(1+\epsilon V)}(d\tau+
\epsilon^{3/2} A)^2
\ee
depending on a harmonic function $V$ and a one--form $A$ on $\R^3$ such that $\nabla V=\nabla\wedge A$ admits a  Newtonian limit
$(h=\lim_{\epsilon\rightarrow 0}(g^{-1}), \theta=d\tau, \nabla=\lim_{\epsilon\rightarrow 0}
\nabla_{{g_\epsilon}})$ with the unparametrised 
geodesic equations of $\nabla$ given by
$d^2{\bf {x}}/d\tau^2=-(1/2)\nabla V$. 
In Section \ref{new_spin} we shall introduce the spinor calculus adopted
to a 3+1 splitting, leading to Newtonian spinors. In Section 
\ref{twistor_sect} we shall review the relativistic twistor theory, and proceed to construct its non-relativistic limit. This will be done by
realising the relativistic twistor space as an affine line bundle
over the total space of the bundle $\OO(2)\rightarrow \CP^1$, 
taking a limit of the resulting patching matrix
(Section  \ref{sect_jump_lines}) and then 
(Section \ref{nrte}) by directly analysing the incidence relation 
for a complexified Minkowski space $M_\C$ with a speed of light 
$c\neq 0$. We shall prove the following
\begin{theo}
\label{theo_1}
Let $PT_c$ be a one-parameter family of rank-two holomorphic vector bundles over $\CP^1$ determined by a patching matrix
\[
F_c=\left (
\begin{array}{cc}
1 & -{(c\lambda)}^{-1} \\
0 & {\lambda^{-2}}
\end{array}
\right ), \qquad \mbox{where}\quad \lambda\in\CP^1.
\]
\begin{enumerate}
\item If $c$ is finite, then the holomorphic sections
of $PT_c\rightarrow\CP^1$ have normal bundle $\OO(1)\oplus\OO(1)$.
The four--dimensional moduli space $M_\C$ of these sections carries a flat conformal  structure such that two points in $M_\C$ are null separated iff the corresponding sections intersect at one point in $PT_c$.
\item If $c=\infty$ then the holomorphic sections
of $PT_c\rightarrow\CP^1$ have normal bundle $\OO\oplus\OO(2)$.
The four--dimensional moduli space $M_\C$ of these sections
admits a fibration $M_\C\rightarrow \C$ defined by a closed one--form
$\theta\in\Lambda^{1}(M_\C)$, and a contravariant metric $h$
of rank $3$ which is non--degenerate on $\mbox{ker}(\theta)$.
\item For any $c\neq 0$ (finite or not) there exists an involution
$\sigma:PT_c\rightarrow PT_c$ which restricts to an antipodal map 
on each section, and such that the $\sigma$--invariant sections
form a real four-manifold $M$ with structures as in $(1)$ and $(2)$.
\end{enumerate}
\end{theo}
This Theorem will be established in
Section  \ref{sect_jump_lines}. We shall explicitly construct
the sections of $PT_c$ which will allow us to identify $c$ with the speed of light. If $c=\infty$ then $F_\infty=\mbox{diag}(1, \lambda^{-2})$ and $PT_\infty=\OO\oplus\OO(2)$, but for finite $c$ there exist holomorphic splitting matrices 
$H$ and $\widetilde{H}$
such that
$F=\widetilde{H}\; \mbox{diag}(\lambda^{-1}, \lambda^{-1})\; H^{-1}$ and
$PT_c=\OO(1)\oplus \OO(1)$. For all non--zero $c$ there exists
a holomorphic fibration $PT_c\rightarrow \OO(2)$. This fibration is trivial if $c=\infty$, which gives rise to  a global 
twistor function on $PT_\infty$ and consequently to a closed
one--form $\theta$ on $M_\C$. The conformal freedom in 
$\theta$ and $h$ are both fixed by a holomorphic canonical one--form 
on the base of the fibration $PT_\infty\rightarrow\CP^1$.

In Section  \ref{sec_kodaira} we shall discuss the Kodaira deformation
theory of the complex structure underlying the non-relativistic twistor space. We shall exhibit a class of deformations which do not preserve the type of the normal bundle of twistor curves, and lead to Gibbons--Hawking metrics. This is a consequence of the non-vanishing of the obstruction group
$H^1(\CP^1, \mbox{End}(N))=\C$, where $N=\OO\oplus\OO(2)$ is the normal bundle to non--relativistic twistor lines.
We shall also discuss deformations for which the twistor curves do not change their holomorphic type. 

In Section \ref{sec_merkulov} we shall use 
Merkulov's relative deformation theory to give a
construction of Newtonian connections on $M$ with no Coriolis term. We shall establish
\begin{theo}
\label{theo_3}
There is a 1-1 correspondence between line bundles over
$PT_\infty$  which are trivial on $\sigma$--invariant sections of $PT_\infty\rightarrow\CP^1$, and Newtonian connections on $M$.
\end{theo}
It may appear that all Theorem \ref{theo_3} does is to reinterpret  the Penrose transform between $\sigma$--invariant sections of $H^1(PT_\infty, \OO)$ and solutions to the Laplace equation on $\R^3$. In fact we are saying much more than that. 
 The Newtonian connections of Theorem \ref{theo_3} will arise
as a family of morphisms
\be
\label{mer_map_i}
N_{\mathcal{F}}\otimes\left(\odot^{2}N_{\mathcal{F}}^{*}\right)
\rightarrow
TM_\C\otimes\mbox{Sym}^2(T^*M_\C),
\ee
where 
$N_{\mathcal{F}}$ is the normal bundle 
to the correspondence space 
\[
{\mathcal F}=\{(p, \xi)\in M_\C\times PT_{\infty}, \xi\in L_p\}\subset M_\C\times 
PT_{\infty}
\]
in the product manifold $M_\C\times PT_{\infty}$ and
where $L_p=\CP^1$ is a rational curve in $PT_\infty$ corresponding to a point
$p$ in the complexified Newtonian space--time $M_\C$.
The contribution from the $H^0(\CP^1, \OO(2))=\C^3$ factor in 
(\ref{mer_map_i}) corresponds (by the Serre duality) to a zero-rest-mass field on $M_\C$ arising from a gravitational potential $V$ in the Gibbons--Hawking family
(\ref{GH_intro}) with $\epsilon=c^{-2}$.

 In  Section \ref{sparling_section} rank--two vector bundles which restrict to $\OO\oplus\OO(2)$ on twistor curves will lead to connections with non--zero Coriolis term. 

 In  Appendix 1 we shall exhibit and interpret a contour integral formula
\[
\psi(x, y, u)=\frac{1}{2\pi i}\oint_{\Gamma\subset \CP^1} e^{-\frac{1}{2}mi(x-iy)\lambda}
g(x+iy+\lambda u, \lambda)d\lambda,
\]
for solutions to the 2+1 Schrodinger equation
$2m\p_u\psi=i({\p_x}^2+{\p_y}^2)\psi$.
Finally in Appendix 
2 we shall show how the spin connection in the Nonlinear Graviton construction arises as a Ward transform of a rank--two holomorphic vector bundle over the relativistic deformed twistor space.

\vspace{2ex}{\bf Acknowledgements.} We are grateful to Christian Duval, George Sparling and Paul Tod for helpful discussions. This work started when MD was visiting the Institute for Fundamental Sciences (IMP) in Tehran in April 2010. MD is grateful
to IMP for the extended hospitality  when volcanic eruption in Iceland
halted air travel in Europe. 
The work of JG has been supported by an STFC studentship.

\section{Newton--Cartan gravity}
The trajectories of test particles in Newtonian physics with a chosen universal
time $t$ are integral curves ${\bf x}={\bf x}(t)$  of the system of ODEs
\be
\label{geodesics_V}
\frac{d^2{\bf x}}{d t^2}=-\nabla V,
\ee
where $V:\R^3\rightarrow \R$ is the Newtonian potential. 
One normally interprets these equations as giving rise to curved paths
in a fixed three-dimensional space. 
The geometric perspective of Cartan \cite{cartan} differs in that the particle trajectories  are instead  regarded as geodesics of some connection in a four-dimensional
space time $M$. To introduce the necessary structures consider a four--manifold $M$ equipped 
with a triple $(h, \theta, \nabla)$,
where $h$ is a degenerate contravariant metric of signature $(0, +, +, +)$,
$\theta$ is a one--form which belongs to a kernel of $h$ (when viewed a
map from $T^*M$ to $TM$), and $\nabla$ is a torsion--free 
affine connection on $M$ with covariant derivative $\nabla$ which
keeps $h$ and $\theta$ parallel. In local coordinates
\be
\label{compatibility}
h=h^{ab}\p_a\otimes\p_b, \quad \theta=\theta_ad x^a, \quad\mbox{and}\quad
\nabla_a h^{bc}=\nabla_a\theta_b=0.
\ee
The last two conditions do  not specify $\nabla$ uniquely.
The most general connection satisfying (\ref{compatibility}) is 
parametrised by an otherwise irrelevant  choice of time--like vector 
$U$ such that
$\theta_aU^a=1$  and a choice of  two--form $F$. The most general connection
such that $U^a\nabla_a U^b=0, \nabla^{[a} U^{b]}=0$ is given by
the Christoffel symbols
\be
\label{g_connection}
\Gamma_{ab}^c=\frac{1}{2}h^{dc}(\p_a h_{bd}+\p_bh_{ad}-\p_d h_{ab})+\p_{(a}\theta_{b)}U^c +\theta_{(a}F_{b)d}h^{dc},
\ee
where $\p_a=\p/\p x^a$, and
$h_{ab}$ is a degenerate metric uniquely determined  by the conditions
\[
h_{ab}U^b=0, \qquad h_{ac}h^{bc}=\delta^b_a-\theta_a U^b.
\]
Connections of the form (\ref{g_connection}) are called {\em Galilean}.
A Galilean connection is called {\em Newtonian} if the two-form $F$ is 
closed, or equivalently if the Trautman condition \cite{Trautman}
\be
\label{trautman}
h^{a[b}{R^{c]}}_{{}{(de)a}}=0
\ee
holds. Here $R^{a}_{bcd}$ is the curvature of $\nabla$.
The condition (\ref{trautman}) is required for consistency with
a non-relativistic limit of general relativity (see
Section \ref{nr_limit}). Even this additional requirement allows
connections which are more general than one needs to reproduce
the Newtonian theory (see Section \ref{corio_ex}).
For a given $(h, \theta)$ the pair $(U^a, F_{ab})$ is defined up to gauge transformations
\[
U^a\rightarrow U^a+h^{ab}\Psi_b, \quad F\rightarrow F+d\Phi,
\]
where $\Psi=\Psi_adx^a$ is an arbitrary one--form, and 
\[
\Phi=\Psi-\Big(U\hook \Psi+\frac{1}{2}h(\Psi, \Psi)\Big)\theta.
\]
Thus, if $\nabla$ is a Newtonian connection (so that $dF=0$), then one 
can  locally set $F=0$ and instead work with a non--trivial vector field $U$.
This approach is adapted in \cite{son2}.

The absence of torsion of $\nabla$, and the consistency 
conditions (\ref{compatibility}) imply the existence of 
a function $t:M\rightarrow \R$ such that $\theta=dt$.
The three--dimensional distribution spanning the kernel of $\theta$ is integrable in the sense of the Frobenius theorem, and 
the manifold $M$ admits a fibration over a real line $\R$
called the time axis, with a coordinate $t$
\[
M\rightarrow M/\mbox{ker}\;(\theta)=\R.
\]
Thus the space--time $M$ in the Newton--Cartan theory is a fibre bundle over
a universal time axis equipped with a Riemannian metric $h$ on the fibers.

The Einstein field equations are
\be
\label{field_eq}
R_{ab}=4\pi G\;\rho \;\theta_a\theta_b,
\ee
where $R_{ab}$ is the Ricci tensor of $\nabla$, the function 
$\rho:M\rightarrow \R$ is the mass density and $G$ is the gravitational 
constant. These equations in particular imply that
the metric $h$ on three-dimensional spatial fibres is flat. This allows an introduction of the {\em Galilean coordinates} $x^a=(t, {\bf x})$ such that
(in general indices $a, b, \dots$ run from 0 to 3 and $i, j, \dots$ run 
from 1 to 3),
\[
h=\delta^{ij}\frac{\p}{\p x^i}\otimes\frac{\p}{\p x^j}, \quad
\mbox{and}\quad \theta=dt.
\]

 Let us  spell out some physical consequences of the 
Newton--Cartan formalism. 
In general, the free falling particles follow the geodesics of $\nabla$ which however
do not need to be affinely parametrised.
The statement that two events are 
simultaneous in Newtonian physics is invariant and coordinate independent.
This is a consequence of the existence of the one--form $\theta$ and the resulting fibration $M\rightarrow \R$. On the other hand one can not say
that two events occur at the same point in 3-space but at different times.
To be able to compare two non-simultaneous events one needs to make an additional choice of Galilean coordinates such that the spatial part of the Newton-Cartan connection vanishes. The existence of such coordinates is a consequence
of
$\nabla h=0$, and the flatness of $h$ which itself follows from the field
equations (\ref{field_eq}). Moreover the distance between two events is only invariantly defined if these events are simultaneous.
\subsection{Example 1. Newtonian connection}
\label{newto_ex}
Take $h$ to be a flat metric on the spatial fibres, i. e. 
$h^{ij}=\delta^{ij}$,
$\theta=d t$ and consider $\Gamma^{i}_{00}=\delta^{ij}\p_j V$ with all other components
of $\nabla$ equal to zero. In this case  
\[F=dt\wedge dV,
\]
and the geodesics of $\nabla$ parametrised by $t$ 
are the integral curves of the Newton equations of motion 
(\ref{geodesics_V}).
The coordinate transformation
\be
\label{gali}
{\bf x}\longrightarrow\ \hat{{\bf x}}= R(t){\bf x}+{\bf a}(t),
\ee
where ${R}$ is an element of ${SO}(3)$ which is allowed to depend
on $t$, gives rise to the  `Coriolis forces' given by the connection components
$
{{\hat\Gamma}^i}_{0j}=\dot{R}_{kj}R^{ik}.
$
A combination of an inertial force $\ddot{\bf a}$
and a centrifugal force $\ddot{R} \bf x$ also appear as components of the connection as
\[
{{\hat\Gamma}^i}_{00}=\delta^{ij}\p_j V+R^i_{k}(\ddot{R}^k_lx^l-\ddot{a}^{k}).
\]
The form (\ref{geodesics_V}) of the geodesic equations is preserved only if
$\dot{R}=0, \ddot{\bf a}=0$. 
\subsection{Example 2. Newton-Cartan connection with the Coriolis force}
\label{corio_ex}
The Newtonian connection of the previous subsection can be modified to a Galilean connection which includes the general
space-time dependent 
Coriolis force ${\bf{B}}={\bf B}({\bf x}, t)$.
In this case the closed two-form $F$ is 
\[
F=-dt\wedge E+\frac{1}{2}F_{ij}dx^i\wedge dx^j, \quad F_{ij}=2\epsilon_{ijk} B^k,
\]
and the corresponding  non-zero connection components are
\be
\label{cor_con}
\Gamma_{ab}^c= h^{cd}\theta_{(a}F_{b)d}, \quad
\Gamma_{00}^i=-E^i, \quad \Gamma_{0j}^i=\frac{1}{2}{F_{j}}^i,
\ee
where $\theta=dt$, $h^{ab}$ is a degenerate metric of signature
$(0, 1, 1, 1)$ such that $h^{ab}\theta_b=0$ which in our case
is chosen to be $\mbox{diag}(0, 1, 1, 1)$.

The field equations together with the Trautman
condition imply ${\bf B}=\nabla W$, for some function $W$, and
\be
\label{new_ricci}
\nabla\cdot {\bf E}=2 |\nabla W|^2, \quad \nabla^2 W=0.
\ee
At this stage we can not assume that ${\bf E}=-\nabla V$.
This can be either put in by hand, or derived from an asymptotic condition that 
${\bf B}\rightarrow{\bf B}(t)$ at spatial infinity. This condition, together with 
the uniqueness theorem for the Laplace equation implies that 
${\bf B}$ can be set to zero. This is because if $x^i\rightarrow {R^i}_j(t)x+a^i(t)$, then
$F_{ij}\rightarrow {R^k}_i {R^l}_j F_{kl}+{R^k}_i\dot{R}_{kj}$, which can be set to zero by a choice of an orthogonal matrix $R(t)$.
Alternatively, the closure of $F$ implies that ${\bf E}$ arises from a potential if ${\bf B}$ does not 
depend on $t$.
If ${\bf E}=-\nabla V$, then 
\[
\nabla^2 W=0, \quad \nabla^{2}(V+W^2)=0,
\]
so that the connection is determined by two harmonic functions.
In Section (\ref{sparling_section}) we shall show how two independent 
harmonic functions arise from holomorphic vector bundles over 
a Newtonian twistor space $PT_{\infty}$ which restrict to a non-trivial vector bundle $\OO\oplus\OO(2)$ on each twistor line.
\section{Newton--Cartan geometry as a limit of General Relativity}
\label{nr_limit}
In this Section we shall show how
Newton--Cartan geometry arises as a degenerate limit of 
General Relativity \cite{kunzle, dautcourt}. 
Let $(M, g(\epsilon))$ be a family of pseudo-Riemannian manifolds
parametrised by $\epsilon>0$, and such that
\[
g^{ab}(\epsilon)=h^{ab}+\epsilon k^{ab}+O(\epsilon^2)
\]
where $h^{ab}$ is a contravariant tensor  on $M$ of signature 
$(0, +, +, +)$. The following result has been 
established in \cite{kunzle}
\begin{theo}
\label{theo_kunz}
Let $\theta=\theta_a dx^a$ be an $\epsilon$--independent 
 one--form such that $h^{ab}\theta_a=0$, normalised by
$k^{ab}\theta_a\theta_b=-1$. 
If the one--form $\theta$ is closed, then the Levi-Civita connection of $g(\epsilon)$ has a well defined limit $\nabla$ as $\epsilon$ goes to 
zero.  It is a Newton-Cartan connection (\ref{g_connection}) with $dF=0$.
\end{theo}
 The Newton--Cartan structure resulting from Theorem \ref{theo_kunz}
is $(h, \nabla, \theta)$. The Ricci tensor
of $g(\epsilon)$ also has a well defined limit, and choosing a perfect
fluid energy-momentum tensor for the relativistic theory leads
(in the limit) to the field equations (\ref{field_eq}). See \cite{dautcourt} for details of this derivation.
\vskip5pt
Another way to proceed \cite{Ehlers} is to define a one-parameter
family of structures
$\{h^{ab}, t_{ab}, \nabla\}$ such that $(t, h)$ are parallel
with respect to $\nabla$
and $t_{ab}h^{bc}=-\epsilon{\delta^c}_a$. For $\epsilon>0$ this gives
pseudo-Riemannian geometry with $g_{ab}=-t_{ab}/\epsilon$, and for $\epsilon=0$
the Newton--Cartan theory with $t_{ab}=\nabla_a t\nabla_b t$, and $\theta_a=\nabla_a t$.
\subsection{Examples}
We shall now give three examples of Newtonian limits corresponding to particular solutions of Einstein equations. The first example is a minor modification 
from \cite{Ehlers} (where some factors appear to be wrong), and corresponds to 
a non--zero Coriolis force and a Newton--Cartan structure of the type 
(\ref{cor_con}). 
The remaining two examples give limits
of anti-self-dual solutions 
and thus correspond to analytic continuation of the Newton--Cartan theory. 
The resulting geodesic equations can be analytically continued to Newton's equations of motion despite the fact that ASD gravitational instantons
in Riemannian signature, or ASD $pp$--waves in neutral signature do not admit 
Lorentzian analytic continuations. Our motivation for considering the limits
of ASD solutions will become clear in Section \ref{sec_ward} when we construct
non--relativistic twistor spaces from a limiting procedure.
\subsubsection{Newtonian limit of Taub-NUT}
Consider a  one-parameter family \cite{Ehlers}  of Lorentzian Taub-NUT metrics, with the mass parameter $m$, and the NUT charge $a$ 
\[
g=\frac{1}{U} dr^2+(r^2+\epsilon a^2)(d\theta^2+\sin^2{\theta}d\phi^2)-
\frac{U}{\epsilon}(dt+2\epsilon a\cos{\theta}d\phi)^2,
\]
where 
\[
U=1-\frac{2\epsilon(mr+a^2)}{r^2+\epsilon a^2}.
\]
Employing the Cartesian  coordinates $(x, y, z)$
\[
r=\sqrt{x^2+y^2+z^2}, \quad \theta=\arccos\frac{z}{\sqrt{x^2+y^2+z^2}},
\quad \phi=\arctan{\frac{y}{x}}
\]
yields the gravitational acceleration
\[
\Gamma_{00}^i=x^i\Big(\frac{2a^2}{r^4}+\frac{m}{r^3}\Big)=\delta^{ij}
\frac{\p}{\p x^i} V, \quad\mbox{where} \quad V= \Big( -\frac{m}{r}-\frac{a^2}{r^2}\Big),
\]
and the Coriolis force
\[
{\Gamma_{0j}}^i={\Gamma_{j0}}^i ={\epsilon^i}_{kj}  B^k,
\quad \mbox{where}\quad B^k=\frac{ax^k}{r^3}=
\delta^{ki}\frac{\p}{\p x^i}\Big(-\frac{a}{r}\Big).
\]
Both forces arise from scalar potentials $V$ and $W=-a/r$, but only $W$ 
is a harmonic function. The two potentials satisfy
\[
\nabla^2 V+2|\nabla W|^2=0
\]
which implies the Ricci-flat condition (\ref{new_ricci}).
\subsubsection{Newtonian limit of gravitational instantons}
\label{GH_limit}
Let $V$ and $A$ be respectively a harmonic function, and a one--form 
on $\R^3$ which satisfy the Abelian monopole equation
\[
dV=*dA,
\]
where $*$ is the Hodge endomorphism of the flat metric on $\R^3$.
The one--parameter family of  Gibbons Hawking metrics \cite{GH} 
\be
\label{gh_earlier}
g=(1+\epsilon V)(dx^2+dy^2+dz^2)+\frac{1}{\epsilon(1+\epsilon V)}(d\tau+
\epsilon^{3/2} A)^2
\ee
has anti--self--dual Weyl tensor and is Ricci--flat. Conversely,
any Riemannian ASD Ricci--flat (and so hyper-K\"ahler) four-manifold 
admitting a  Killing vector preserving 
the  hyper-Kahler structure is of the form
(\ref{gh_earlier}). The Newtonian limit of (\ref{gh_earlier}) is
\be
\label{GHcon}
h^{ij}=\delta^{ij},\quad {\Gamma^{i}}_{\tau\tau}=
\frac{1}{2}\delta^{ij}\frac{\p V}{\p x^j}, \quad \theta=d\tau,
\ee
where all other components of $\Gamma^{a}_{bc}$ vanish.
\vskip5pt
For example $V=1/r$ corresponds to the ASD Taub-NUT metric. It is
an asymptotically locally flat (ALF) gravitational instanton.
The metric (\ref{gh_earlier}) is everywhere regular, and 
asymptotically approaches the non--trivial $S^1$ fibration over $S^2$
with the monopole number $1$. In the Newtonian limit the fibration becomes trivial, so that the connection is asymptotically flat, but the genuine singularity arises at $r=0$.

\subsubsection{Newtonian limit of anti-self-dual plane waves}
The most general ASD Ricci-flat metric admitting a parallel Killing vector is of the form \cite{plebanski}
\be
\label{pp_wave_m}
g=dwdx+dzdy+\gamma(w, y)dw^2.
\ee
If the coordinates $(w, z, x, y)$ on $M$, and the arbitrary function
$\gamma=\gamma(w, y)$ are taken to be real, then the metric has signature $(2, 2)$.
Set
\[
w=\frac{t}{\sqrt{\epsilon}}+u, \quad x= \frac{t}{\sqrt{\epsilon}}-u
\]
and redefine $\gamma$. This, for a given $\gamma$, gives
a one--parameter family of metrics 
\[
g=\frac{1}{\epsilon} dt^2-du^2+dzdy+\gamma(t+\sqrt{\epsilon}u, y)
(dt+\sqrt{\epsilon}du)^2.
\]
with a Newtonian limit
\be
\label{newtonian_pp}
h\equiv \mbox{lim}_{\epsilon\rightarrow 0}
(g^{-1})=-\p_u\odot\p_u+2\p_y\odot\p_z, \quad
\theta\odot \theta\equiv \mbox{lim}_{\epsilon\rightarrow 0}
(\epsilon g)=dt\odot dt, \quad
\Gamma_{tt}^z=-\frac{\p \gamma}{\p y},
\ee
where now $\gamma=\gamma(t, y)$. This structure can, for certain $\gamma$s, be analytically continued to a
Newtonian theory\footnote{Consider a different way of taking the limit, and set
$
z={t}/{\sqrt{\epsilon}}+u,  y= {t}/{\sqrt{\epsilon}}-u
$
so that
\[
g={\epsilon}^{-1} dt^2+dwdx-du^2+\gamma(w, t-\sqrt{\epsilon} u)dw^2.
\]
Now the limit is
\[
h=-4\gamma\p_x\odot\p_x+\p_x\odot\p_w-\p_u\odot\p_u,
\quad 
\Gamma_{ww}^x=\frac{\p \gamma}{\p w}, \quad
\Gamma_{tw}^x=\frac{\p \gamma}{\p t}.
\]
This appears to be different than (\ref{newtonian_pp}), but isn't,
as the coordinate transformation
$
\hat{x}=x+4\kappa,  \hat{w}=w, {u}=u,  \hat{t}=t$ 
where $\gamma=\p_w \kappa$
gives
$
h=\p_{\hat{w}}\odot\p_{\hat{x}}-\p_{\hat{u}}\odot
\p_{\hat{u}}.
$}, where the signature of $h$ is $(0, 1, 1, 1)$. 
\section{Newtonian spinors}
\label{new_spin}
Let $(M, g)$ be a (pseudo) Riemannian four--manifold. Locally there exist
complex rank-two vector bundles $\spp$ and $\spp'$ over $M$ equipped with covariantly constant symplectic structures $\ve$ and $\ve'$
parallel w.r.t the Levi--Civita connection and such that 
\be
\label{can_bun_iso}
\C\otimes T M\cong {\spp}\otimes {\spp'}
\ee
is a  canonical bundle isomorphism, and
\[
g(p_1\otimes q_1,p_2\otimes q_2)
=\ve(p_1,p_2)\ve'(q_1, q_2)
\]
for $p_1, p_2\in \Gamma(\spp)$ and $q_1, q_2\in \Gamma(\spp')$.
We use the conventions of Penrose and Rindler  \cite{PR}, where the spinor indices are capital letters, unprimed for sections of $\spp$ and primed for sections of $\spp'$. For example $p^{A}$ denotes a section of $\spp$, 
and $q_{A'}$ a section of $(\spp')^*$.
The symplectic structures $\ve_{AB}$
and $\ve_{A'B'}$ (such that
$\ve_{01}=\ve_{0'1'}=1$) are used to lower and
raise the spinor indices according to $p_{A}:=p^{B}\ve_{BA}, 
p^A=\ve^{AB}p_B$. The properties of the complex conjugation on spinors
depends on the signature of $g$.
\begin{itemize}
\item If $g$ is Lorentzian, then the spinor complex conjugation
is a map $\spp\rightarrow\spp'$ given by
\[
p^A=(p^0, p^1) \rightarrow \ov{p}^{A'}=(\ov{p^0}, \ov{p^1}).
\]
Linear $SL(2, \C)$ transformations on $\spp$ induce Lorentz 
transformations on vectors, and $SO(3, 1)=SL(2, \C)/\Z_2$.
\item If $g$ is Riemannian, then the complex conjugation preserves
the type of spinors, i.e. it maps sections of $\spp$ to sections of $\spp$
and sections of $\spp'$ to sections of $\spp'$. It is given by
\be
\label{riem_rel}
p^A=(p^0, p^1)\rightarrow \hat{p}^A=(\ov{p^1}, -\ov{p^0}), 
\quad
q_{A'}=(q_{0'}, q_{1'})\rightarrow \hat{q}_{A'}=(-\ov{q_{1'}}, \ov{q_{0'}}).
\ee
In the Riemannian case the structure group is not simple, and
$SO(4, \R)=SU(2)\times SU(2)'/\Z_2$, where the spin groups 
$SU(2)$ and $SU(2)'$ act linearly on sections of $\spp$ and $\spp'$ 
respectively.
\end{itemize}
Now assume that $g$ has Lorentzian signature, and chose a time--like 
unit vector $T$ to perform the 3+1 split of $g$. This vector gives a map
from $\spp$ to $\spp'$ given by \cite{som} 
\be
\label{31spinors}
p^A\rightarrow {T_{A}}^{A'}p^A,
\ee
where we have chosen
\be
\label{unit_vector}
T^{AA'}=\frac{1}{\sqrt{2}}(o^{A}o^{A'}+\iota^{A}\iota^{A'})
\ee
where $(o^{A'}, \iota^{A'})$ is the normalised basis
of $\spp'$, i.e. $\ve_{A'B'}o^{A'}\iota^{B'}=1$, and
$(o^{A}, \iota^{A})$ is the normalised basis of $\spp$.
Under this map the isomorphism (\ref{can_bun_iso}) becomes
\begin{eqnarray}
\label{3spinor}
\C\otimes TM&\cong& \spp'\otimes \spp'\nonumber\\
&=&(\spp'\odot \spp')\oplus \Lambda^2(\spp'),
\end{eqnarray}
where $\Lambda^2(\spp')$ is a rank--one vector bundle,
and $\mbox{Sym}^2(\spp')$ is a rank--three vector bundle 
isomorphic to a bundle of self-dual two-forms
on $M$.
 
The isomorphism (\ref{3spinor}) gives
the orthogonal decomposition:
\be
\label{decomposition}
V^{AA'}\rightarrow V^{AA'}{T_{A}}^{B'}=V^{(A'B')}+ \frac{1}{2}g(V,T)\varepsilon^{A'B'}.
\ee
The symmetric matrix $V^{(A'B')}$ represents a three-vector in the  
space
$t=$const and the multiple of $\varepsilon^{A'B'}$ is the component in normal direction.

Consider  the flat metric
\begin{eqnarray}
\label{metric_md}
g&=&\ve_{AB}\ve_{A'B'}dx^{AA'}\otimes dx^{BB'}\\
&=& c^2dt^2-dx^2-dy^2-dz^2,\nonumber
\end{eqnarray}
where
\be
\label{point_matrix}
x^{AA'}= 
\frac{1}{\sqrt{2}}
\left(\begin{array}{cc}
ct+z& x+iy\\
x-iy&ct-z
\end{array}\right)
\ee
(so that $T^a\p/\p x^a=c^{-1}\p/\p t$).
The map (\ref{31spinors}) gives
\[
x^{AA'}{T_{A}}^{B'}=x^{A'B'}-\frac{1}{2}\ve^{A'B'}ct,
\]
where
\be
\label{three_matrix}
x^{A'B'}= 
\frac{1}{{2}}
\left(\begin{array}{cc}
x-iy& -z\\
-z&-x-iy
\end{array}\right).
\ee
In our discussion of the Newtonian limit of general relativity in Section
\ref{nr_limit} we needed to single out a time--like one--form $\theta_a=\nabla_a t$  in the kernel
of the three--metric $h$. We shall use the dual vector--field
to perform the 3+1 
decomposition (\ref{decomposition}),  and construct the 
isomorphism (\ref{31spinors}). Therefore the two--component spinor 
formalism for
Newton--Cartan structures involves only one type of spinors
(which we choose to be sections of $\spp'$).

It is natural to ask whether the Newton--Cartan connection on $TM$ 
is induced from some spin connection on $\spp'\rightarrow M$. The answer to this question appears to be `no'
which is related to the fact that the Galilean boosts do not have a double 
cover.  A Newton--Cartan connection takes its 
values  in the Lie algebra of the Galilean group. However we find that
the homogeneous Galilean
transformations \[
({\bf x}, t)\rightarrow (R{\bf x}+{\bf v}t, t)
\]
can not be realised spinorially as
\be
\label{gal_spin}
{Y^{A'}}_{B'}\rightarrow {M^{A'}}_{C'}\; {Y^{C'}}_{D'}\; {N^{D'}}_{B'}
\ee
where ${Y^{A'}}_{B'}=x^{AA'}T_{AB'}$
or $Y=Y_0+ct{\bf I}$, and $Y_0$ is trace-less.
The condition (\ref{gal_spin}) forces $M=N^{-1}$ which then implies ${\bf v}=0$.
The choice of the unit vector (\ref{unit_vector})  breaks the Lorentz invariance.
A subgroup of $SL(2, \C)$ which acts on $\spp$ and $\spp'$, and preserves (\ref{unit_vector}) is  $SU(2)$. Its
action on spinors only generates rotations, and not Lorentz boosts. The more general 
transformations of the form $Y\rightarrow MYN +PY$ do not work either.

It is nevertheless possible to construct a spin connection in the Newtonian 
limit, although to  achieve a non--zero result the speed of light
has to be regarded as a part of the constant metric on the fibres
of $T^*M$ in the Cartan formalism. To see it
consider the post--Newtonian limit of GR given by
\be
\label{post_new}
g=-c^2e^{2V/c^2} dt^2+\delta_{ij}dx^idx^j.
\ee
Employing the spin frame $(ce^{V/c^2}dt, dx^i)$ yields a spin connection which 
vanishes in the limit $c\rightarrow \infty$. Let us instead consider
\[
\eta_{ab}=\left(
\begin{array}{cccc}
-c^2 & 0&0&0 \\
0&1&0&0\\
0&0&1&0\\
0&0&0&1
\end{array}
\right), \quad e^0=e^{V/c^2}dt, \quad e^i=dx^i
\]
so that (\ref{post_new}) is given by $g=\eta_{ab}e^ae^b$.
Then the Cartan equations $de^a+{\omega^a}_b\wedge e^b=0$ together with a limit
$c\rightarrow\infty$ yield the non--vanishing connection given by
$
{\omega^i}_0=(\p^i V)\theta.
$
\section{Twistor theory}
\label{twistor_sect}
\subsection{Review of relativistic twistor theory}
In this Section we shall recall basic facts about twistor theory, 
following \cite{PR} and \cite{MD}.

The theory was put forward by Roger Penrose 
\cite{Penrose_twistor_alg} in the late 1960s.
The primary  motivation was 
to unify general relativity and quantum theory in a non--local theory 
based on complex numbers, but 
a good starting point for our discussion 
is Penrose's contour integral formula for 
wave equation in Minkowski space \cite{Pen_2, EPW} 
\be
\label{penrose_form}
\phi({x,y,z,t})=\frac{1}{2\pi i}\oint_{\Gamma\subset \CP^1}
f({{(x+iy)}}+{{(ct+z)}}\lambda,{{(ct-z)}} 
+{{(x-iy)}}\lambda, \lambda)d\lambda,
\ee
where $\Gamma\subset \CP^1$ is a closed contour and
the function $f$ is holomorphic on an intersection of two open sets in 
$\CP^1$ away from  poles inside $\Gamma$.
Differentiating inside the integral shows that
\be
\label{wave_eq}
\frac{1}{c^2}\frac{\p^2 \phi}{\p t^2}-\frac{\p^2 \phi}{\p x^2}-
\frac{\p^2 \phi}{\p y^2}-\frac{\p^2 \phi}{\p z^2}=0.
\ee
The formula (\ref{penrose_form}) 
gives real  solutions to the wave equation
in Minkowski space from holomorphic functions of three variables,
and the next step is to recognise these variables as coordinates in some complex
three--dimensional manifold.

The five dimensional space ${PN}$ of light rays
in the Minkowski space is a hyper--surface in a three-dimensional complex
manifold ${PT}=\CP^3-\CP^1$ called the projective twistor 
space. 
Let $(\omega^0, \omega^1, \pi_{0'}, \pi_{1'})\sim  \gamma (\omega^0, \omega^1, \pi_{0'}, \pi_{1'})    , \gamma\in \C^*$ 
with $(\pi_{0'}, \pi_{1'})\neq (0, 0)$ be 
homogeneous coordinates of a twistor. 
The Minkowski space and the twistor space are linked by
the incidence relation
\be
\label{4d_incidence}
\left (
\begin{array}{cc}
\omega^0\\
\omega^1 
\end{array}
\right )
=
\frac{1}{\sqrt{2}}
\left (
\begin{array}{cc}
ct+z & x+iy\\
x-iy & ct-z
\end{array}
\right )
\left (
\begin{array}{cc}
\pi_{0'} \\
\pi_{1'} 
\end{array}
\right )
\ee
where $x^\mu=(t, x, y, z)$ are coordinates of a point in Minkowski space.
It then follows that if two points in  Minkowski space
are incident with the same twistor, then they are separated by a null geodesic.
The spinor form of the incidence formula above is
\be
\label{Twistor_equation}
\omega^{A}=x^{AA'}\pi_{A'},
\ee
where the displacement vector $x^{AA'}$ is given by (\ref{point_matrix}).

Let $Z=(\omega^A, \pi_{A'})$ represent homogeneous coordinates of a twistor.
The non--projective twistor space $\C^4-\C^2$
is equipped with a Hermitian inner product 
\be
\label{inner_null}
\Sigma(Z, \bar{Z})=\omega^{A}\bar{\pi}_{A}+\bar{\omega}^{A'}\pi_{A'}.
\ee
The signature of
$\Sigma$ is $(+ + - -)$ and endomorphisms of 
${\C^4}$ preserving $\Sigma$  as well as the orientation of $\C^4$
form a group $SU(2, 2)$ which is
locally isomorphic to the conformal group of
the Minkowski space. The projective twistor space is divided into three regions
depending on the sign of $\Sigma(Z, \bar{Z})$. The hypersurface
\be
\label{pnspace}
{PN}=\{(\omega^A, \pi_{A'})\in {PT}, \Sigma(Z, \bar{Z})=0 \}\subset {PT}
\ee
is preserved by the conformal transformations of the Minkowski space.

Fixing the coordinates $(x, y, z, t)$ of a  space--time point 
$p$
in (\ref{4d_incidence}) gives a projective line $L_p=\CP^1$ in ${PT}$.
If the coordinates of $p$ are real, then $L_p$ is contained in the
hypersurface ${PN}$. Conversely, fixing a point 
in ${PN}$ gives a light--ray in the Minkowski space.

Points in ${PT/PN}$ can be interpreted in terms of the complexified Minkowski space
$M_\C=\C^4$ where they correspond to null two--dimensional planes
($\alpha$--planes) with self--dual tangent bi-vector. This, again, is a direct consequence of (\ref{4d_incidence}) where now the coordinates $x^{AA'}$ 
are complex (Figure 1).
\begin{center}
\includegraphics[width=8cm,height=3cm,angle=0]{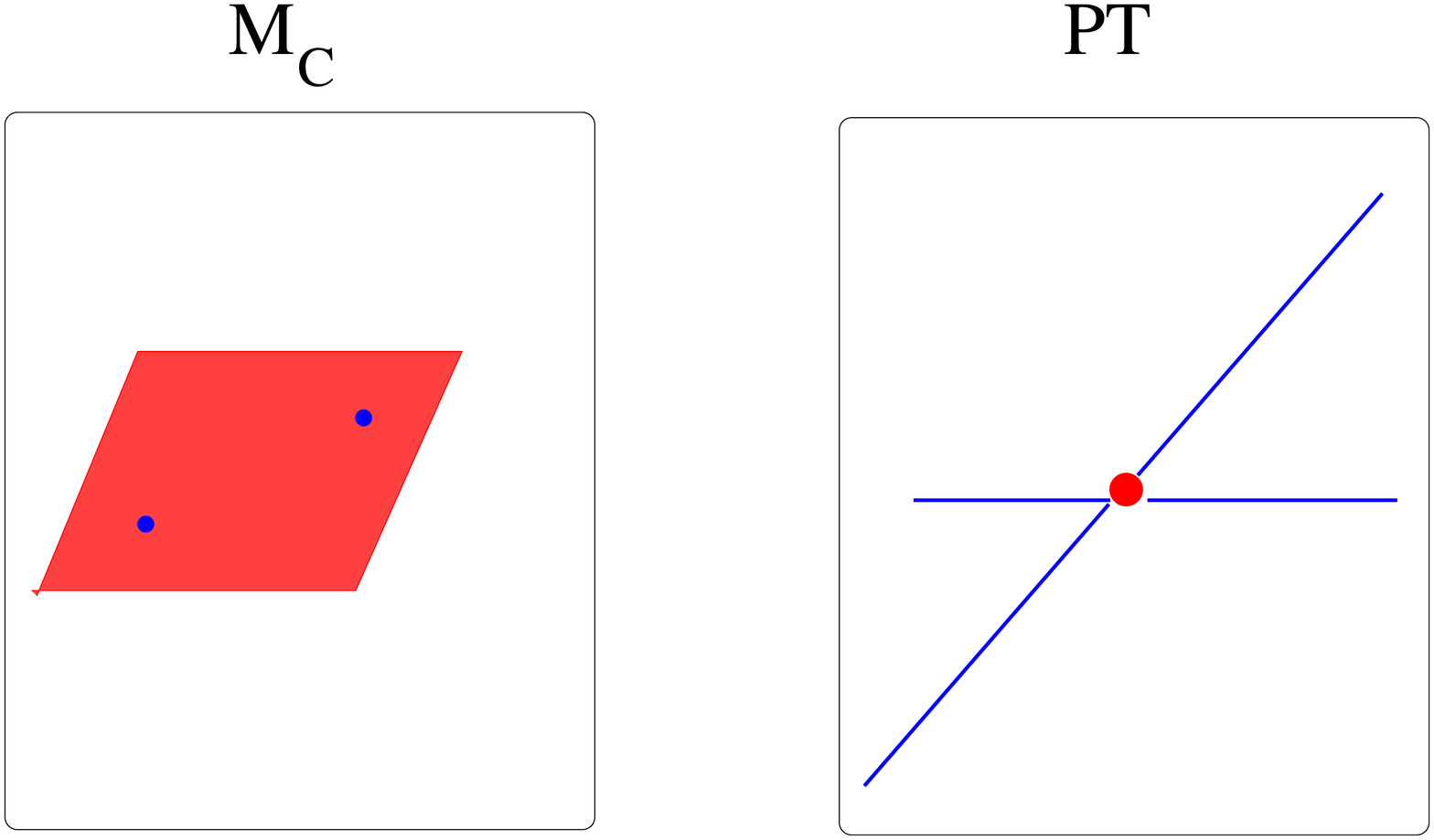}
\begin{center}
{{\bf Figure 1.} {\em Twistor relativistic incidence relation. 
Two points on an $\alpha$--plane in $M_\C$ correspond to two
twistor lines passing through a point in $PT$.}}
\end{center}
\end{center}
\vskip5pt
There is a $\CP^1$ worth of $\alpha$--planes through each point of $M_\C$.
For any $\pi_{A'}\in\CP^1$ a corresponding $\alpha$--plane is spanned by a
two-dimensional twistor distribution
\be
\nabla_{A}=\pi^{A'}\p_{AA'},\quad \mbox{where}\quad\p_{AA'}=
\frac{\p}{\p x^{AA'}}.
\label{alpha_dist}
\ee

Finally we can give a twistor interpretation
of the contour integral formula (\ref{penrose_form}). Consider a function
$f=f(\omega^A/\pi_{1'}, \pi_{0'}/\pi_{1'})$  
which is holomorphic on an intersection 
of two open sets covering $PT$ (one of these sets is defined 
by $\pi_{1'}\neq 0$ and the other by $\pi_{0'}\neq 0$). Restrict $f$ to a 
rational curve $L_p\subset{PN}$ and integrate $f$ along 
a contour in $L_p$. This yields (\ref{penrose_form}) with 
$\lambda =\pi_{0'}/\pi_{1'}$.  The function $f$, when regarded as defined 
on the non--projective twistor space, must be homogeneous of degree
$-2$. 
\subsection{Non-relativistic twistor theory}
Our construction of a non-relativistic twistor space for the flat Newtonian 
space time  will proceed in two steps. We shall first represent the 
relativistic twistor space introduced in the last Section as an affine line bundle over
$\OO(2)$, and show that in the limit $c\rightarrow\infty$ the normal bundle
of sections of $PT\rightarrow \CP^1$ jumps from $\OO(1)\oplus\OO(1)$ to 
$\OO\oplus\OO(2)$. This will be done is Section \ref{sect_jump_lines}. 
In Section \ref{nrte} we shall relate the jumping lines approach
to a non-relativistic limit of the 3+1 split of the twistor incidence relation.
\subsubsection{Proof of Theorem \ref{theo_1}}
\label{sect_jump_lines}
Let $[\pi_{0'}, \pi_{1'}]$ be homogeneous coordinates on
$\CP^1$. Cover $\CP^1$ with two open sets
\be
\label{UUopen_sets}
U=\{[\pi]\in\CP^1, \pi_{1'}\neq 0\}, \quad
\widetilde{U}=\{[\pi]\in\CP^1, \pi_{0'}\neq 0\}
\ee
and set $\lambda=\pi_{0'}/\pi_{1'}$ on $U\cap\widetilde{U}$. 
The Birkhoff-Grothendieck theorem  states that any rank-$k$ holomorphic vector bundle
over $\CP^1$ is isomorphic to a direct sum of line bundles
\[\OO(m_1)\oplus\OO(m_2)\oplus\dots\oplus \OO(m_k)
\]
for some integers $m_1, m_2, \dots, m_k$. Moreover the transition matrix
$F:\C^*\rightarrow GL(k, \C)$ of this bundle can be written
as
\be
\label{BH_split}
F=\widetilde{H}\; \mbox{diag}(\lambda^{-m_1}, \lambda^{-m_2}, \dots, \lambda^{-m_k})\; H^{-1},
\ee
where $H:U\rightarrow GL(k, \C)$ and $\widetilde{H}:\widetilde{U}\rightarrow
GL(k, \C)$ are holomorphic.

Let $PT_c\longrightarrow \CP^1$ be
a one-parameter family of
rank-two vector bundles determined by a patching matrix
\[
F_c=\left (
\begin{array}{cc}
1 & -{(c\lambda)}^{-1} \\
0 & {\lambda^{-2}}
\end{array}
\right ),
\]
where $c$ is a constant. If $c=\infty$
then $F_\infty=\mbox{diag}(1, \lambda^{-2})$
is the patching matrix for $PT_{\infty}=\OO\oplus\OO(2)$ with $H$ and $\widetilde{H}$ in (\ref{BH_split}) both equal to the identity matrix. 
From now on we shall identify the constant $c$ with the speed of light, and 
refer to $PT_{\infty}$ as the Newtonian twistor space. 

 If $c\neq \infty$ then
\[
F_c=\left(
\begin{array}{cc}
0 & -{c}^{-1} \\
c & {\lambda}^{-1}
\end{array}
\right) 
\left(
\begin{array}{cc}
{\lambda}^{-1} & 0 \\
0 & {\lambda}^{-1}
\end{array}
\right ) 
\left(
\begin{array}{cc}
1 & 0 \\
-c\lambda & 1
\end{array}
\right )
\]
which is of the form (\ref{BH_split}) with
\[
\widetilde{H}=\left(
\begin{array}{cc}
0 & -{c}^{-1} \\
c & {\lambda}^{-1}
\end{array}
\right),\quad
H=\left(
\begin{array}{cc}
1 & 0 \\
-c\lambda & 1
\end{array}
\right )^{-1}.
\]
Thus $PT_c=\OO(1)\oplus\OO(1)$ if the speed of light is
non-zero and finite. 
This is the relativistic twistor space, with the holomorphic sections
of $PT_c\rightarrow \CP^1$ parametrised by points in the complexified Minkowski space $M_\C$.
\vskip5pt
Let $x^{AA'}$ given by (\ref{point_matrix})
be a displacement vector of a point in $M_\C$ from the origin,
and let $T^{AA'}$
be a unit vector (\ref{unit_vector})
with respect to the flat metric
(\ref{metric_md}).
Set
\[
\omega={T_A}^{A'}\omega^A\pi_{A'}=\frac{1}{\sqrt{2}}
(\omega^1\pi_{0'}-\omega^{0}\pi_{1'})
\]
and define inhomogeneous coordinates $(Q, T)$ and $(\widetilde{Q}, \widetilde{T})$
in pre-images
of $U$ and $\widetilde{U}$ in $PT_c$ by
\be
\label{patch_U}
\Big(Q=2\frac{\omega}{{\pi_{1'}}^2}, 
T=  \frac{\sqrt{2}}{c}\frac{\omega^1}{\pi_{1'}}\Big) \quad \mbox{on}\quad U
\ee
and
\[
\Big(\widetilde{Q}=2\frac{\omega}{{\pi_{0'}}^2}, 
\widetilde{T}=
\frac{\sqrt{2}}{c}
\frac{\omega^0}{\pi_{0'}}\Big) \quad \mbox{on}\quad \widetilde{U}.
\]
On the pre-image of $U\cap\widetilde{U}$ in $PT_c$ we have
\begin{eqnarray}
\label{patch_flat}
\widetilde{Q}&=&\Big(\frac{\pi_{1'}}{\pi_{0'}}\Big)^2 Q=
\frac{1}{\lambda^2}Q\\
\widetilde{T}&=&\frac{\sqrt{2}\omega^0}{c\pi_{0'}}
-\frac{\sqrt{2}\omega^1}{c\pi_{1'}}
+\frac{\sqrt{2}\omega^1}{c\pi_{1'}}
=T-\frac{2\omega}{c\pi_{0'}\pi_{1'}}=T-\frac{1}{\lambda c}Q,\nonumber
\end{eqnarray}
or
\[
\left (
\begin{array}{cc}
\widetilde{T}\\
\widetilde{Q} 
\end{array}
\right )=F_c \left (
\begin{array}{cc}
{T}\\
{Q} 
\end{array}
\right).
\]
Note that on $PT_\infty$ there exists a global twistor function of weight 0, given by $T=\widetilde{T}$.
\vskip5pt
Restricting the inhomogeneous coordinates to a
twistor line $\omega^{A}=x^{AA'}\pi_{A'}$ gives
\begin{eqnarray}
\label{rest_lines}
Q&=&-(x+iy)-2\lambda z+\lambda^2 (x-iy), \quad
T=t-\frac{1}{c}(z-\lambda(x-iy))\quad \mbox{on}\quad
U\\
\widetilde{Q}&=&(x-iy)-2\tilde{\lambda} z-\tilde{\lambda}^2 (x+iy), \quad
\widetilde{T}=t+\frac{1}{c}(z+\tilde{\lambda}(x+iy))\quad \mbox{on}\quad
\widetilde{U},\nonumber
\end{eqnarray}
where $\tilde{\lambda}=\pi_{1'}/\pi_{0'}$.
Thus, if $c=\infty$, then 
$T=\widetilde{T}=t$ on the twistor lines, where $t$ is the global time coordinate on the complexified Newtonian space-time.

The algebraic geometry of holomorphic sections of $PT_c\rightarrow\CP^1$ determines the metric  
structure on $M_\C$: two points in $M_\C$ are null separated iff the corresponding sections intersect at one point in $PT_c$. Infinitesimally, a vector in $T_pM_\C$ is null
if the corresponding section of $N(L_p)$ vanishes at one point. This condition is equivalent to the existence of the unique solution 
$\lambda=\lambda_0$ for a simultaneous system 
\[
\delta Q=0, \quad \delta T=0,
\]
where $Q$ and $T$ are given by (\ref{rest_lines}), and the variation $\delta$ is w.r.t the moduli $(x, y, z, t)$ and not $\lambda$.
If $c$ is finite, then the conformal structure on $M_\C$ is determined by $\mbox{det}(dx^{AA'})=0$, and the conformal factor is fixed by a two-form $dQ\wedge dT$ on the fibres of $PT_c$. If $c=\infty$, then the
the conformal structure on Minkowski space degenerates in the following way:
The equation $\delta{T}=0$ implies that $t=\mbox{const}$, and in the Newtonian limit the null separation implies simultaneity of events. We also deduce the existence of a canonical closed one--form
$dT$ on $PT_\infty$ which yields $\theta\equiv dt$ on $M_\C$.  This is the {\em clock} of
Newton-Cartan theory which defines a fibration
of $M_\C$ over the time axis
\be
\label{fibration_over_C}
M_\C\rightarrow M_\C/{\mbox{ker}(\theta)}=\C. 
\ee
The section of $N(L_p)$ will have a single zero if the two roots of the remaining equation $\delta Q=0$ coincide. Evaluating the discriminant of the quadratic form  leads to a degenerate conformal structure on $M_\C$, which is
determined by $h=dx^2+dy^2+dz^2$ on three--dimensional  fibres of (\ref{fibration_over_C}).
The conformal factor is fixed by the canonical $\OO(2)$--valued one-form  
$\ve^{A'B'}\pi_{A'}d\pi_{B'}$ on the fibres
of $PT_{\infty}\rightarrow \CP^1$ by
$h^{ab}=\ve^{A'(C'}\ve^{D')B'}$. The relation between $h_{ab}$ and $h^{ab}$ is
provided by a unique choice of a vector field $U$ 
such that  $U\hook \theta=1$, and $h_{ab}U^a=0$. This gives a degenerate
contravariant metric on $T^*M_\C$ such that $h^{ab}\theta_a=0$.
The discussion of the intrinsic definition of the connection is postponed 
to Section \ref{sec_merkulov}.
\vskip5pt
 Now we shall move on to discuss the reality 
conditions and establish point (3) in Theorem \ref{theo_1}. 
The Euclidean real slice of the complexified Minkowski space $M_\C$
is characterised by an involution
$\sigma:PT_c\rightarrow PT_c$ given by
\be
\label{o2inv}
\sigma(Q, \lambda, T)=\Big(-\frac{\ov{Q}}{\ov{\lambda}^2}, 
-\frac{1}{\ov{\lambda}}, -\ov{T}+\frac{1}{c}\frac{\ov{Q}}{\ov{\lambda}}\Big)
\ee
so that $\sigma^2=\mbox{Id}$ on the projective twistor space. 
This involution has no fixed points, because on the non--projective
twistor space it squares to minus identity, and so for any $[Z]\in PT_c$
there is a unique line joining $[Z]$ to $[\sigma[Z]]$. These are
the {\em real twistor curves}, which correspond to points
in the Euclidean slice of $M_\C$.  These $\sigma$--invariant sections
are characterised by
\[
\sigma{(Q)}=-(x+iy)+\frac{2z}{\ov{\lambda}}+\frac{x-iy}{\ov{\lambda}^2}, \quad Q=-(x+iy)-2\lambda z+\lambda^2 (x-iy)
\]
so that $(x, y, z)$ are real. Similarly 
\[
\sigma{(T)}=t-\frac{1}{c}(z+\frac{1}{\ov{\lambda}}(x-iy)), \quad T=t-\frac{1}{c}(z-\lambda(x-iy))
\]
so that $\ov{t}=-t$, and $t=i\tau$ is purely imaginary. Thus the real 
Newtonian twistor lines in $PT_\infty$ are
\be
\label{newton_lines}
Q=-(x+iy)-2\lambda z+\lambda^2 (x-iy), \quad
\widetilde{Q}=\lambda^{-2}Q, \quad T=\widetilde{T}=i\tau
\ee
where $(x, y, z, \tau)$ are real coordinates on $M$.\koniec

It is worth remarking that in the limit when $c\rightarrow \infty$
the real (i.e. $\sigma$--invariant) twistor curves 
lie on the hypersurface (\ref{pnspace}). To see it consider the 
inner product $\Sigma_c:=c^{-1}\Sigma$, where $\Sigma$ is  given by
(\ref{inner_null}). Then
\[
\Sigma_{\infty}(Z, \bar{Z})=(T+\bar{T})(|\pi_{0'}|^2+|\pi_{1'}|^2).
\]
Thus the Newtonian twistor space is divided into two regions
$\mbox{Re}(T)>0$ and $\mbox{Re}(T)<0$ separated by the 
five--dimensional space $PN_\infty$ of null twistors $\mbox{Re}(T)=0$. The difference 
between the Lorentzian reality conditions (twistor curves contained in $PN_c$)
and Riemannian reality conditions (twistor curves invariant under an
anti-holomorphic involution) disappears in the Newtonian limit $c\rightarrow\infty$.

\vskip5pt
To sum up,  we have exhibited a realisation of the 
relativistic twistor space
$\OO(1)\oplus\OO(1)$
as an affine line bundle over the total space
of the line bundle $\OO(2)\rightarrow\CP^1$. In the limit $c=\infty$ the holomorphic structure of $PT_c$
changes discontinuously: The total space of $PT_c$, and the normal bundle of the twistor curves both jump to $\OO\oplus\OO(2)$. Note that this jumping is a purely 
holomorphic phenomenon: the bundles $PT_c$ are all
the same for any $c$ finite or not from the topological perspective.
Our findings agree with the analysis of \cite{tod} (as well as with 
the recent work \cite{H2}), where the jump from $\OO(1)\oplus\OO(1)$ to 
$\OO\oplus\OO(2)$ was interpreted as a singularity of the conformal 
structure.
\subsubsection{Non--relativistic incidence relation}
\label{nrte}
Combining the $(3+1)$ splitting (\ref{three_matrix})
with the incidence relation (\ref{Twistor_equation})
yields
\be
\label{n_twistor}
x^{A'B'}\pi_{B'}-\frac{1}{2}ct \pi^{A'}=\omega^{A'} 
\ee
where $x^{A'B'}$ is given by (\ref{three_matrix}), and
$\omega^{A'}={T_A}^{A'}\omega^A$. This equation blows up in the
limit $c\rightarrow\infty$ but
the twistor function $\omega=\omega^A {T_{A}}^{B'}\pi_{B'}$ remains finite.
If a pair $(x^{A'B'}_0, t_0)$ solves
the twistor equation (\ref{n_twistor}), 
then so does
\[
x^{A'B'} = x^{A'B'}_0+\pi^{A'}\rho^{B'}+\rho^{A'}\pi^{B'}, \quad t= t_0-\frac{2}{c}\rho_{C'}\pi^{C'}
\]
for any constant spinor $\rho^{A'}\in \Gamma(\spp')$.
The image of
the two dimensional twistor distribution (\ref{alpha_dist}) tangent to $\alpha$-planes
under (\ref{decomposition}) is
\[
{T^{A}}_{A'}:\nabla_A\rightarrow \nabla_{A'}=\pi^{B'}\p_{A'B'}+\frac{1}{2c}\pi_{A'}
\frac{\p}{\p t}.
\]
In the Newtonian limit this gives an integrable  2-plane distribution
(the mini-twistor distribution)
 on the constant time
three-dimensional fibres of $M\rightarrow \C$
\[
{\mathcal{D}}=\mbox{span}\{\p_x+i\p_y+\lambda\p_z, \p_z-\lambda(\p_x-i\p_y)\}.
\]
The non-relativistic twistor space $PT_{\infty}$ is the three-dimensional space
of leaves of ${\mathcal{D}}$ in the correspondence space
$M_{\C}\times\CP^1$. To specify a leaf pick a point on the time axis, and  choose a two plane (a point in a mini-twistor space $\OO(2)$) which is null with
respect to the holomorphic metric $dx^2+dy^2+dz^2$ on the fibres of 
(\ref{fibration_over_C}). Thus, in the Newtonian limit the $\alpha$--planes 
become space-like, and 
they lie on the fibres of (\ref{fibration_over_C}) (Figure 2).
\begin{center}
\includegraphics[width=7cm,height=7cm,angle=0]{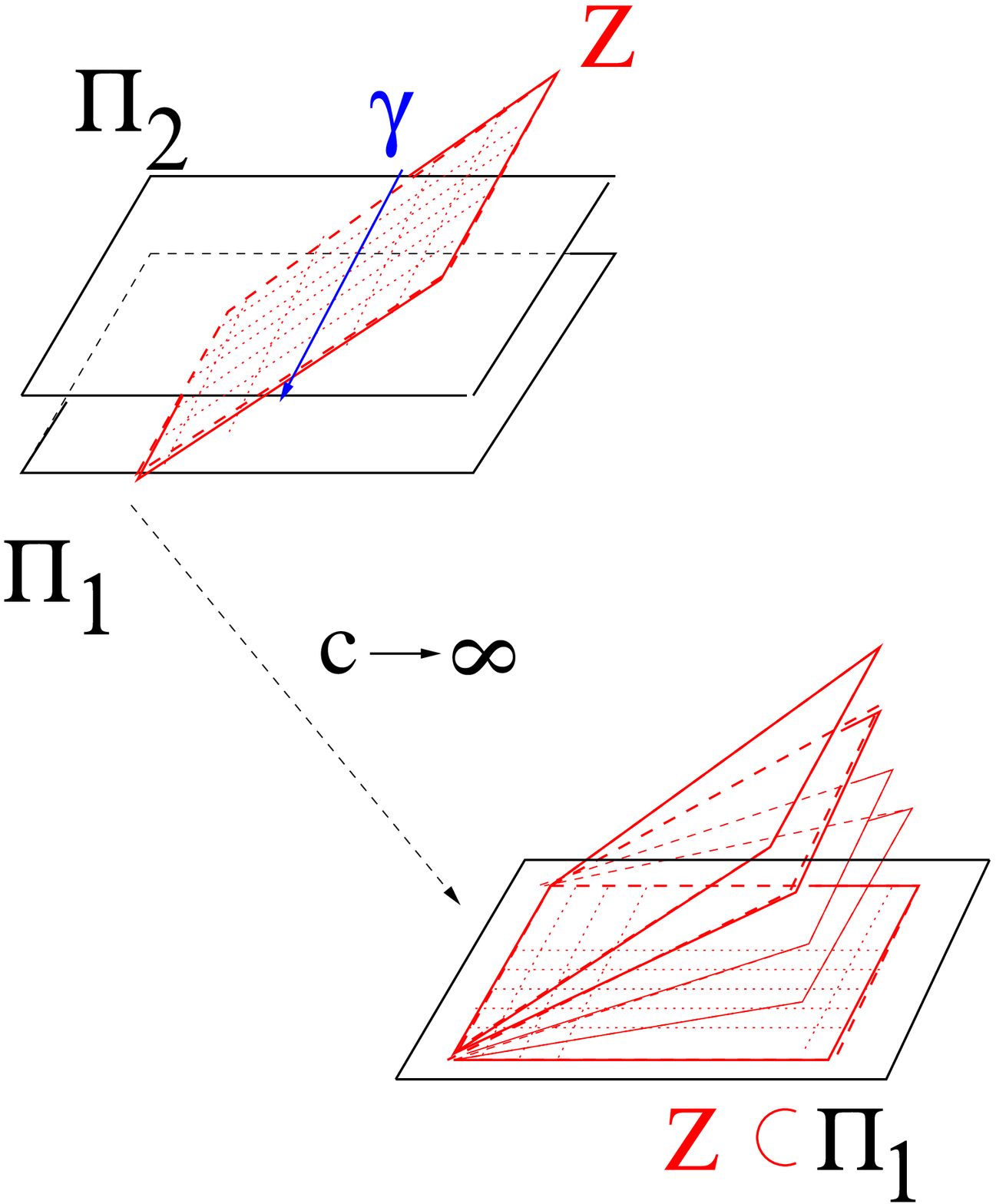}
\begin{center}
{{\bf Figure 2.} {\em An $\alpha$--plane $Z$ intersects hyper-planes $\Pi$
of constant time in $M_\C$ in null lines $\gamma$. 
In the non--relativistic limit  $Z$ becomes a subspace of a constant-time hyper-plane $\Pi_1$.}}
\end{center}
\end{center}
A point $p\in M_\C$ gives a time coordinate 
and a $\CP^1$ worth of 2-planes on the three--dimensional spatial fiber. 
This is a rational curve $L_p=\CP^1$ in $PT_{\infty}=\OO\oplus\OO(2)$ 
lying in a fibre $T=const$.
The normal bundle $N(L_p)$ of a rational curve $L_p$ corresponding to
$p\in M$ is ${\mathcal O}\oplus {\mathcal O}(2)$ as we have established in Section \ref{sect_jump_lines}. This can also be seen from
the exact sequence
\[
0\rightarrow \spp'\otimes{\mathcal{O}}(-1)\rightarrow \C^4=\C^3\oplus\C \rightarrow N(L_p)\rightarrow 0
\]
as the last map is, in the spinor notation, given by 
\[(V^{(A'B')}, V^{A'B'}\ve_{A'B'})\rightarrow (V^{A'B'}\pi_{A'}\pi_{B'}, 
V^{A'B'}\ve_{A'B'} )
\]
clearly projecting onto ${\mathcal O}(2)\oplus {\mathcal O}$,
and the decomposition of $\C^4$ into $\C^3\oplus \C$ is
given by (\ref{decomposition}).

 This is a complexified
picture. If the reality conditions (\ref{riem_rel}) 
and (\ref{newton_lines})
are imposed,
then the intersection of a complex $\alpha$-plane
with a real $\tau=const$ slice is a real straight line - a geodesic
of the spatial metric - with a tangent
vector $\pi^{(A'}\hat{\pi}^{B')}$ (Figure 3).
\vskip5pt
\begin{center}
\includegraphics[width=8cm,height=5cm,angle=0]{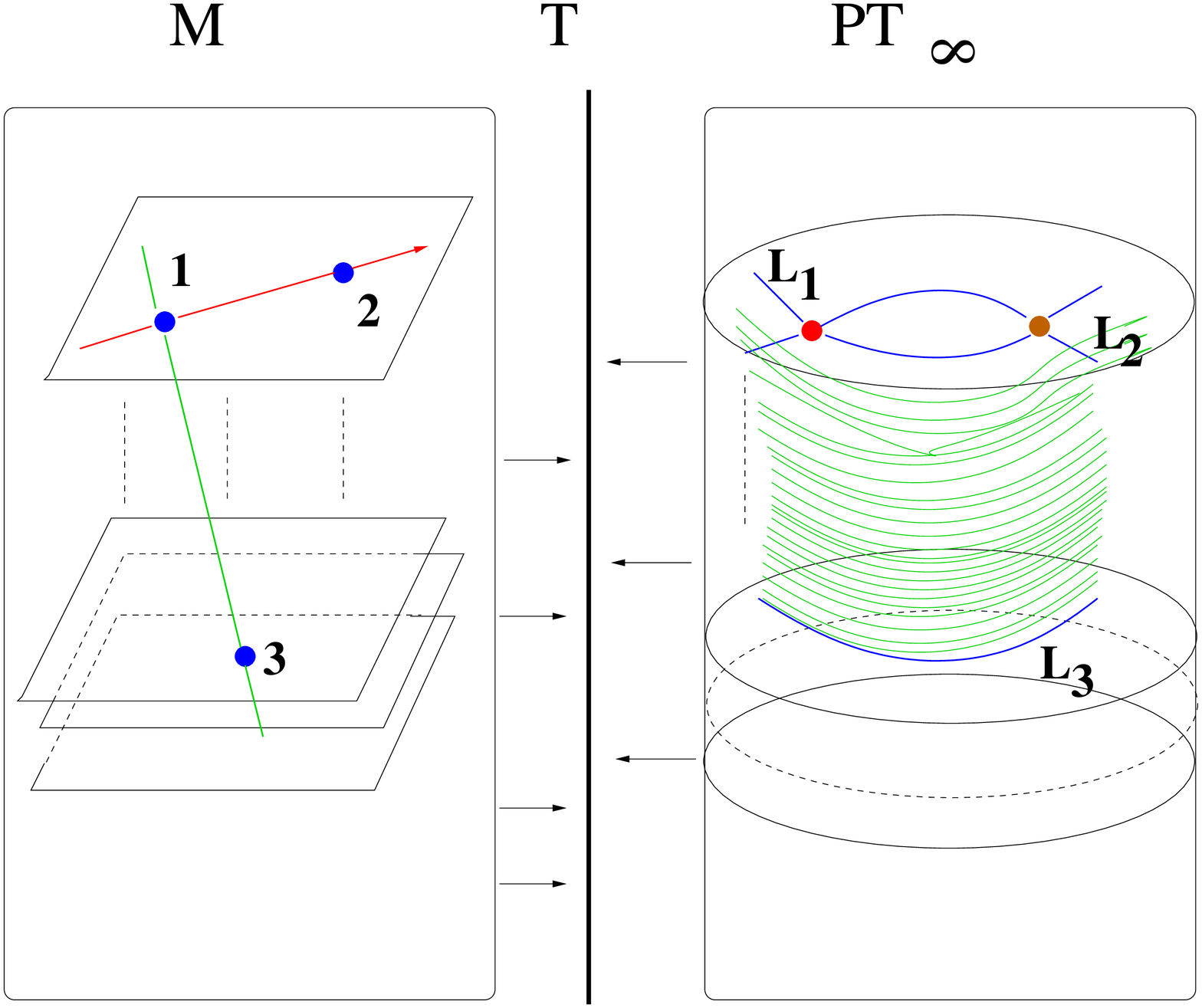}
\begin{center}
{{\bf Figure 3.} {\em A real section of the non--relativistic 
incidence relation. Oriented straight lines in the fibres of
$M\rightarrow \R$ correspond to points in $PT_{\infty}$. Two simultaneous
events ${\bf 1}$ and  ${\bf 2}$ in Newtonian space--time correspond rational curves
$L_1, L_2$ intersecting at two points in $PT_\infty$. A time--like geodesic
between two non--simultaneous events ${\bf 1}$ and ${\bf 3}$ corresponds to 
a ruled surface in $PT_\infty$.}}
\end{center}
\end{center}
\vskip5pt
The twistor curves invariant under the involution 
(\ref{o2inv}) correspond to points in the 
flat Newtonian space-time $M=\R^4$. The points in $PT_\infty$
 correspond to oriented straight lines on the fibres of $M\rightarrow \R$. Thus two real curves $L_1$ and $L_2$ in $PT_{\infty}$ either do not intersect, or intersect at two points $Z$ and $\sigma(Z)$ which correspond to a straight 
line (with two possible orientations) joining $p_1$ and $p_2$ in $M$.


\section{Kodaira deformations}
\label{sec_kodaira}
A way to introduce the curvature on $M_\C$ in relativistic
twistor theory is to deform the complex structure of $PT_c$. The Kodaira 
theorems \cite{kodaira} guarantee that, under the additional assumption of 
the deformation being `small', the deformed twistor space will still
admit a four-dimensional moduli space $M_\C$ of rational curves
with normal bundle $\OO(1)\oplus\OO(1)$. The Nonlinear Graviton theorem of 
Penrose \cite{penrose} then implies that $M_\C$ admits a holomorphic conformal structure
with anti-self-dual (ASD) Weyl tensor, and moreover that all
ASD conformal structures correspond to some deformed twistor spaces. 
If the deformation preserves the fibration of $PT_c$ over $\CP^1$, and the 
symplectic two-form $\ve_{AB}d\omega^A\wedge d\omega^B$ on the fibres
of this fibration, then the resulting conformal structure contains a 
Ricci-flat metric. 

In this section we shall show that the Kodaira deformation theory, when applied
to a Newtonian twistor space $PT_{\infty}$, in general leads to the jumping lines
phenomenon: the deformed twistor curves have normal bundle 
$\OO(1)\oplus\OO(1)$, and thus $M_\C$ carries a non--degenerate metric,
rather than a Newton--Cartan structure. We can phrase it by saying that
the Newtonian space--times are unstable under the general Kodaira 
deformations, but we should note that this instability is a purely holomorphic 
feature of the underlying twistor space, and not a dynamical process.
\subsection{ Jumping lines and Gibbons--Hawking metrics} 
\label{instability}
To consider a sub-class of deformations we shall use our construction
(Section \ref{sect_jump_lines}) of the 
undeformed Newtonian twistor space $PT_{\infty}$ as a total space of 
a trivial holomorphic line bundle
over a total space of $\OO(2)\rightarrow\CP^1$
\[
\OO\oplus\OO(2)=\C\times\OO(2)
\]
where sections of $PT_{\infty}\rightarrow\OO(2)$ restricted to sections of
$\OO(2)\rightarrow\CP^1$ are twistor curves.
This motivates a  replacing the undeformed  twistor space $PT_{\infty}$ by
\[
L\rightarrow \OO(2),
\]
where $L$ is a non-trivial line bundle becoming trivial on each twistor line. 
To construct  $L$ as a deformation of $\OO\oplus \OO(2)$ 
consider the deformed patching relations 
\be
\label{GH_defo}
\widetilde{T}=T+\epsilon f,
\ee
where $f\in H^1(\CP^1, \OO(2))$ represents the cohomology 
class, and $\epsilon$ is a deformation parameter. 
Restrict this cohomology class to a section
of $PT_{\infty}\rightarrow \CP^1$ and pull back to the correspondence space 
so that $f=\tilde{h}-h$,
where $h$ and $\tilde{h}$ are holomorphic in the open sets 
$U$ and $\widetilde{U}$ defined by 
 (\ref{UUopen_sets}). Therefore
\[
T-\epsilon h=\widetilde{T}-\epsilon\tilde{h}=t,
\]
where $t$  does not depend on $\pi$ by the Liouville theorem
and thus can be used as a coordinate on the moduli space of curves in $L$.
The deformed $\CP^1$s (in a patch containing $\lambda=0$) are
\[
\omega(\pi)=\pi_{A'}\pi_{B'} x^{A'B'}, \quad T=t+\epsilon h,
\]
where
\begin{eqnarray}
\label{splitting_h}
h&=&\frac{1}{2\pi i}\oint_{\Gamma}\frac{ f(\rho, x)(\pi\cdot \iota)}{(\pi\cdot\rho)(\iota\cdot \rho)}\rho\cdot d\rho, \\
\tilde{h}&=&\frac{1}{2\pi i}\oint_{\widetilde{\Gamma}}\frac{ f(\rho, x)(\pi\cdot \iota)}{(\pi\cdot\rho)(\iota\cdot \rho)}\rho\cdot d\rho. \nonumber
\end{eqnarray}
Here $\iota_{A'}$ is a spinor  (a choice of which arises because of the non-uniqueness of the splitting). The contours $\Gamma$ and $\widetilde{\Gamma}$
are homologous to the equator of the twistor curve $L_p=\CP^1$ in an 
intersection of two open sets covering $L_p$, and such that $\Gamma-\widetilde{\Gamma}$ surrounds the point $\rho_{A'}=\pi_{A'}$ in $L_p$ (Figure 4).
\begin{center}
\includegraphics[width=5cm,height=3cm,angle=0]{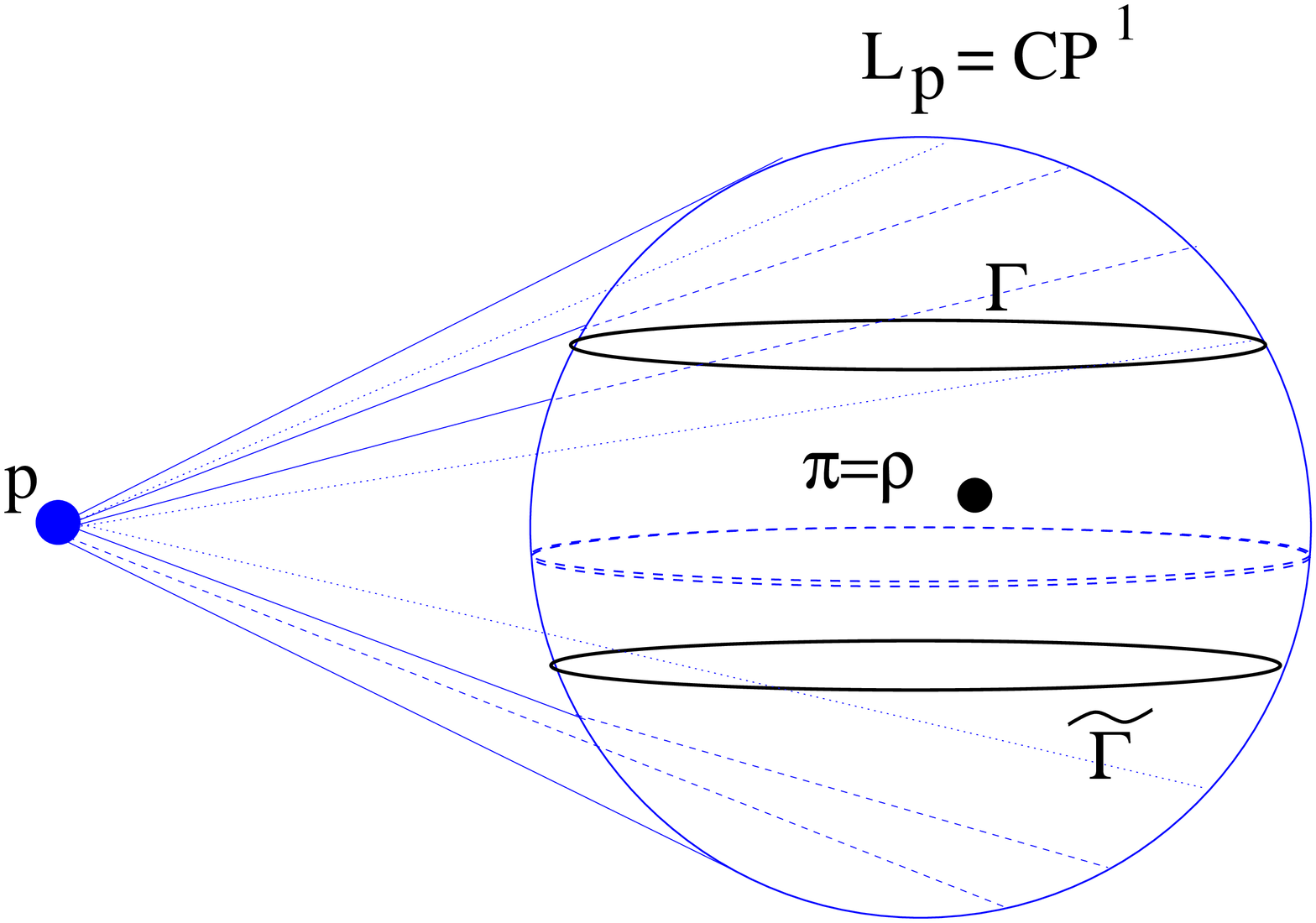}
\begin{center}
{{\bf Figure 4.} {\em Splitting formula. The contour 
$(\Gamma-\widetilde{\Gamma})\subset \CP^1$ surrounds a 
point $\pi_{A'}=\rho_{A'}$.}}
\end{center}
\end{center}
\vskip5pt
The Ward transform \cite{ward} of $L$ gives
\be
\label{ward_1}
\epsilon\nabla_{A'}h=\Phi_{A'B'}\pi^{B'},
\ee
where 
\[
\Phi_{A'B'}=\epsilon A_{A'B'}+\epsilon \varepsilon_{A'B'}V
\]
is given by
\be
\label{twistor_monopole}
\Phi_{A'B'}=\frac{\epsilon}{2\pi i}\oint_{\Gamma}\frac{\iota_{A'} \rho_{B'}}{\iota.\rho}
\frac{\p f}{\p \omega}\rho.d\rho
\ee
and the one form $A_{A'B'}=\Phi_{(A'B')}$ together with a function $V$
satisfy  the monopole equation
\be
\label{monopole}
dV=*dA
\ee
or
$
\nabla_{A'B'}V=\nabla^{C'}_{(A'}A_{B')C'}.
$
Therefore $\nabla_{A'}=\pi^{B'}\p_{A'B'}$ is not a twistor distribution, as it does not Lie-derive $\omega$. The deformed twistor distribution
(also called a Lax pair) is
\begin{eqnarray}
\label{deformed_twist}
L_{A'}&=&\nabla_{A'}-\Phi_{A'B'}\pi^{B'}\frac{\p}{\p t}\\
&=&\pi^{B'}{\bf e}_{A'B'}\nonumber,
\end{eqnarray}
where ${\bf e}_{A'B'}$ are four vector fields on $M_\C$.
This is a jumping line phenomenon: the deformation
changes the normal bundle to $\mathcal{O}(1)\oplus\mathcal{O}(1)$. The vector fields in the Lax pair (\ref{deformed_twist})
\[
{\bf e}_{A'B'}=\p_{A'B'}-\Phi_{A'B'}\p_t=\{\p_{A'B'}-\epsilon 
A_{(A'B')}\p_t, \epsilon V\p_t\}
\]
give rise, by ${\bf e}_{A'B'} {\bf e}^{A'B'}$ to a conformally rescaled Gibbons-Hawking metric. Imposing the reality conditions $t=i\tau$, where 
$\tau\in\R$ and choosing a conformal factor $V$ gives 
the ASD, Ricci--flat metric
\[
g=V (dx^2+dy^2+dz^2)+V^{-1}(\epsilon^{-1} d\tau +A)^2.
\]
Therefore the deformation (\ref{GH_defo}) does not preserve the holomorphic 
type of twistor curves. They jump from the Newtonian $\OO\oplus\OO(2)$ type
to relativistic $\OO(1)\oplus\OO(1)$.

This was to be expected as the general deformations will not preserve the type of the Newtonian normal bundle $N=\mathcal{O}\oplus\mathcal{O}(2)$. This is because $H^1(\CP^1, \mbox{End}(N))\neq 0$. If we regard
a section of $N$ as a column vector with first entry taking values in 
$\mathcal{O}$ and the second entry taking values in 
$\mathcal{O}(2)$, the endomorphisms
of $N$ are represented by two by two matrices
of the form
\be
\label{matrix_b}
\left(\begin{array}{cc}
\alpha& \beta \\
\gamma& \delta
\end{array}\right)
\ee
where $\alpha, \delta\in H^1(\CP^1, \mathcal{O}), \beta\in H^1(\CP^1, \mathcal{O}(-2)), \gamma\in H^1(\CP^1, \mathcal{O}(2))$.
Thus
\[
H^1(\CP^1, \mbox{End}(N))=
\]
\[
H^1(\CP^1, \mathcal{O})\oplus
H^1(\CP^1, \mathcal{O}(-2))
\oplus
H^1(\CP^1, \mathcal{O}(2))
\oplus
H^1(\CP^1, \mathcal{O})=
\C
\]
as the second entry in the sum does not vanish. 
\subsection{A twisted photon construction for Newtonian  
space--times}
\label{tf_def}
The argument above shows that general Kodaira deformations
of $PT_{\infty}$ do not preserve the type of the normal bundle. We have
seen that a deformation which does not preserve the one--form 
$dT$ certainly leads to jumping lines. We can look at the restricted class
of deformations which preserve $dT$, and thus do not change the normal 
bundle. This requires the coefficient $\beta$ in the matrix (\ref{matrix_b})
to vanish. One way to achieve it is to take the twistor space
to be a product $\C\times{\mathcal Z}$, where ${\mathcal Z}$
is a two--dimensional complex manifold with a three--parameter family of curves
with normal bundle $\OO(2)$. Thus  ${\mathcal Z}$ is a mini--twistor space
for a general Einstein--Weyl structure in three dimensions \cite{hitchin}, 
and the corresponding $M$ is a product of an Einstein--Weyl three--manifold
with a line. This is not the structure we are seeking, as the metric
(or in fact a conformal structure) on the three--dimensional 
spatial slices of $M$ will in general 
be curved, whereas we require it to be flat.
In the remainder of this section we shall explore another possibility, and 
deform the relation between a projective and non-projective twistor space in
a way which preserves the one--form $dT$. This is analogous
to the construction of \cite{D99} which in turn was motivated by Ward's
twisted photon construction \cite{ward}. 

The canonical bundle of $PT_{\infty}$ restricted to a twistor line is
$\OO(-4)$. The argument works for $PT_c$ with any non--zero $c$,  as restricting the canonical bundle
$\kappa\rightarrow PT_c$ to a line $L_p$ with a normal bundle  $N(L_p)$
gives
\be
\label{canonical_b}
\kappa|_L=T^*L\otimes \Lambda^2(N(L_p))^*=\mathcal{O}(-4),
\ee
as $T^*L=\mathcal{O}(-2)$.

Let us assume that the line bundle $L=\kappa^{*}\otimes {\mathcal O}(-4)$ is non--trivial on $PT_{\infty}$, but trivial when restricted to twistor lines,
and deform the twistor space in a way which preserves the fibration over $\CP^1$. Therefore
\[
\tilde{\pi}_{A'}=e^f\pi_{A'}, 
\]
where $f=\tilde{h}-h$ is the cohomology class defining $L$ which 
gives rise to the 
abelian monopole 
(\ref{ward_1}). 
This cohomology class coincides with that from Section \ref{instability}
and its splitting is given by
(\ref{splitting_h}). 
The deformed twistor distribution
\be
\label{def_dist}
L_{A'}=\nabla_{A'}-\pi^{B'}\Phi_{A'B'}\Upsilon, \quad \mbox{where}\quad
\Upsilon=\pi^{A'}\frac{\p}{\p \pi^{A'}}
\ee
Lie--derives $e^{-h}\pi_{A'}$.
%
To this end we note that the $\Upsilon$ term can be removed from the Lax pair
by a Mobius transformation of $\pi$ iff $\Phi$ is a gradient:
If $\Phi_{A'B'}=\nabla_{A'B'}U$, then setting
$
\tilde{x}^{A'B'}=x^{A'B'},\tilde{\pi}^{A'}=(\exp{U})\pi^{A'}
$
gives 
\[
L_{A'}=\nabla_{A'}+(\nabla_{A'}U )\;\Upsilon=(\exp{U})\widetilde{\nabla}_{A'}.
\]
In this case the deformed twistor distribution in Frobenius 
integrable for any 
function $U$.  
\section{Newtonian connections from Merkulov's relative deformation theory}
\label{sec_merkulov}
The original Non--Linear Graviton construction of Penrose  \cite{penrose} yields  Ricci--flat ASD metrics
from the holomorphic geometry of associated twistor spaces. In a far reaching generalisation of this 
construction Merkulov \cite{mer, mer2} developed a twistor correspondence between
complete analytic families of complex submanifolds on a given complex manifold, and a class of  torsion--free affine connections
(which he called the $\Lambda$--connections)
on the moduli space of these submanifolds. This has subsequently
led to the celebrated solution of the holonomy problem \cite{mer3}.

In this Section we shall adapt Merkulov's construction to the Newtonian twistor correspondence, and show that
the twistor space $PT_\infty$ introduced  earlier in this paper leads to a family of Newton--Cartan connections
depending on five arbitrary functions. Two of these functions can be set to zero by fixing the overall conformal
factors in the degenerate Newtonian metric $h$ , and the associated clock one--form $\theta$ in the kernel of $h$.
The remaining three functions correspond to the gravitational force. By analysing the
non--relativistic limit of the Gibbons--Hawking twistor space we shall show that the gravitational force
arises from a harmonic gravitational potential which is given by a choice of a holomorphic line--bundle
over $PT_\infty$ trivial on twistor lines. 

\vskip5pt

Let ${J_p}^k$ be an ideal of holomorphic functions on $M_\C$ which vanish to
order $k$ at $p\in M_\C$. Therefore $T_pM_\C=(J_p/{J_p}^2)^*$. The 
second--order tangent bundle $T^{[2]}M_\C$ is defined as a union 
over all points in $M_\C$ of 
second order tangent spaces  $T^{[2]}_pM_\C\equiv (J_p/{J_p}^3)^*$.
Thus  an element of  $(T^{[2]}_pM_\C)^*$ consists of the first two non--zero 
terms of a Taylor  expansion of a function vanishing at $p$, and a section
of $T^{[2]}M_\C$ is a second order linear differential operator
$V^{[2]}=V^a(x)\p_a+V^{ab}(x)\p_a\p_b$. We also have an exact sequence
\[
0\rightarrow T M_\C\rightarrow T^{[2]}M_\C\rightarrow \mbox{Sym}^2(TM_\C)\rightarrow 0
\]
such that $V^a\rightarrow (V^a, 0)$, and $(V^a, V^{ab})\rightarrow V^{ab}$.
A  torsion--free affine connection $\nabla$ 
on $TM_\C$ is then equivalent to a linear map
\be
\label{m_maps}
\gamma: T^{[2]} M_\C\rightarrow TM_\C,
\ee
where 
\[
\gamma (V^{[2]})=(V^a+{\Gamma^a}_{bc}V^{bc})\p_a
\]
for some functions ${\Gamma^a}_{bc}$ which are to be identified with the Christoffel symbols
of $\nabla$.

In \cite{mer,mer2} Merkulov proposed a twistor construction of 
a class of maps (\ref{m_maps}). In some cases
(e.g. the Nonlinear Graviton construction) Merkulov's 
construction leads to a unique connection. We will show that in the case
of the non--relativistic twistor theory a unique  Newtonian connection
in Merkulov's class is provided by a natural choice of a holomorphic
line  bundle over $PT_\infty$.

The general construction, adapted to three--dimensional complex manifolds
$PT_c$ which fiber holomorphically over $\CP^1$, and admit
a moduli space $M_\C$ of holomorphic curves can be summarised as follows.
Let ${\mathcal U}$ and 
$\widetilde{{\mathcal U}}$ be a covering of $PT_c$ by two open sets which is a deformation of the covering (\ref{patch_flat}). Thus $(Q, T, \lambda)$
are holomorphic coordinates in  ${\mathcal U}$ and
$(\widetilde{Q}, \widetilde{T}, \tilde{\lambda})$ are holomorphic
coordinates on  $\widetilde{{\mathcal U}}$. The patching relation
on the overlap ${\mathcal U} \cap \widetilde{{\mathcal U}}$ is
$\tilde{\lambda}=\lambda^{-1}$ and 
$\tilde{w}^A=\tilde{w}^A(w^B, \lambda)$, where
$w^A=(T, Q)$ and $\tilde{w}^A=(\widetilde{T}, \tilde{Q})$.

We shall make use of  the Kodaira isomorphism, which relates
vector fields on $M_\C$ to global sections of the normal bundle $N_{\mathcal{F}}$
to the correspondence space 
\[
{\mathcal F}=\{(p, \xi)\in M_\C\times PT_{c}, \xi\in L_p\}\subset M_\C\times PT_{c}
\]
in the product manifold $M_\C\times PT_{c}$.
The associated double fibration
\[
\label{doublefib}
{M_\C}\stackrel{\alpha}\longleftarrow 
{\mathcal F}\stackrel{\beta}\longrightarrow {PT_{\infty}}.
\]
yields 
\[
N(L_p)=N_{\mathcal{F}}|_{\alpha^{-1}(p)}
\]
where $L_p=\beta\circ\alpha^{-1}(p)=\CP^1$ is the rational curve in $PT_c$ corresponding to a point in
$M_\C$. The patching for this normal bundle is given by
\begin{equation}
\label{merpatch}
F_{\, B}^{A}:=\left[\frac{\partial\tilde{w}^{A}}{\partial w^{B}}\right]_{\mathcal{F}}=\begin{pmatrix}\p_T\widetilde{T} & \p_Q\widetilde{T}\\
\p_{T}\widetilde{Q} & \p_Q\widetilde{Q}
\end{pmatrix}_{\mathcal{F}}
\end{equation}
where the subscript $\mathcal{F}$ means
the  pull-back to the correspondence space. To exhibit
the Kodaira isomorphism take a vector field given by a differential operator  
$V=V^{a}\p_a$ on $\mathcal{F}$, and apply it to the twistor functions 
$\tilde{w}^{A}(x^{a},\lambda)$ to obtain a global section of $N_{\mathcal{F}}$ given by
\begin{equation}
V(\tilde{w}^A)= F_{\, B}^{A} V(w^B) 
\end{equation}
The map (\ref{m_maps})  arises by taking a section of $T^{[2]}M$
and similarly applying the associated differential operator  
to the twistor functions. We find
\begin{equation}
V^{[2]}(\tilde{w}^A)=F_{\, B}^{A} V^{[2]}(w^B)+ 
F_{\, BC}^{A}V^{ab}\partial_{a} w^{B}\partial_{b} w^{C}
\label{eq:obstruction}
\end{equation}
\[
\mbox{where}\quad F_{\, BC}^{A}:=\left[
\frac{\partial^{2}\tilde{w}^{A}}{\partial w^{B}\partial w^{C}}\right]_{\mathcal{F}}.
\]
The term involving $F_{\, BC}^{A}$ is an obstruction to (\ref{eq:obstruction})
representing a global section of $N_{\mathcal{F}}$; only if $F_{\, BC}^{A}$
splits as
\begin{equation}
F_{\, BC}^{A}=-\tilde{\sigma}_{\, EF}^{A}F_{\, B}^{E}F_{\, C}^{F}+F_{\, D}^{A}\sigma_{\, BC}^{D}\label{eq:SplittingProblem}
\end{equation}
for some 0-cochain $\{\sigma, \tilde{\sigma} \}$ of $N_{\mathcal{F}}\otimes\left(\odot^{2}N_{\mathcal{F}}^{*}\right)$
will we arrive at
\[
V^{[2]}(\tilde{w}^A)+
V^{ab}\tilde{\sigma}_{\, EF}^{A}\partial_{a}\tilde{w}^{E}\partial_{b}\tilde{w}^{F} =
F_{\, B}^{A} V^{[2]}(w^A)+
V^{ab}F_{\, B}^{A}\sigma_{\, EF}^{B}\partial_{a}w^{E}\partial_{b}w^{F}.
\]
In solving the splitting problem (\ref{eq:SplittingProblem}) we have
constructed a global section of $N_{\mathcal{F}}$ out of a section
of $T^{[2]}M$, and via the Kodaira isomorphism we have therefore
constructed a vector field out of a section of $T^{[2]}M$. This is
the map (\ref{m_maps}), and we can read off the connection components from
\begin{equation}
\partial_{a}\partial_{b} w^{A}+\sigma_{\, BC}^{A}\partial_{a} w^{B}\partial_{b}w^{C}=\Gamma_{\, ab}^{c}\partial_{c}w^{A}.
\label{ReadOffGamma}
\end{equation}
One can check that $\Gamma_{\, ab}^{c}$ transforms correctly as an
affine connection under a change of coordinates.

 The splitting problem (\ref{eq:SplittingProblem}) is solvable for
any patching in the isomorphism class $N_{\mathcal{F}}$ iff
\be
\label{coho_m}
H^{1}\left(\mathbb{CP}^{1},\, N_{\mathcal{F}}\otimes\left(\odot^{2}N_{\mathcal{F}}^{*}\right)\right)=0.
\ee 
The procedure is however non--unique (and hence leads to a
family of connections) unless 
$H^{0}\left(\mathbb{CP}^{1},\, N_{\mathcal{F}}\otimes\left(\odot^{2}N_{\mathcal{F}}^{*}\right)\right)$ vanishes. Otherwise an element of this cohomology group can be 
added to $\{\sigma, \tilde{\sigma} \}$ . 
In the relativistic case  $N_{\mathcal{F}}|_{L_p}=\mathcal{O}(1)\oplus\mathcal{O}(1)$
which implies that (\ref{coho_m}) holds,
$H^{0}\left(\mathbb{CP}^{1},\, N_{\mathcal{F}}\otimes\left(\odot^{2}N_{\mathcal{F}}^{*}\right)\right)=0$,
and therefore the splitting is unique.
\subsubsection*{Example. Merkulov's connection for ASD plane waves}
To illustrate how this procedure works in practice we will consider the simple
twistor space (originally due to Sparling) described by the infinitesimal deformation of the flat relativistic twistor space $PT_c$ given by
(\ref{ap_ham}), where $f=(\omega^{0})^4/(4\pi_{0'}\pi_{1'})$. 
Integrating the deformation equations and 
introducing the inhomogeneous coordinates on the deformed twistor space by 
$w^A=\omega^A/\pi_{1'}$ on ${\mathcal U}$ and
$\tilde{w}^A=\tilde{\omega}^A/\pi_{0'}$ on $\widetilde{{\mathcal U}}$
yields
\be
\label{defor_s}
\tilde{w}^{0}=\frac{\pi_{1^{\prime}}}{\pi_{0^{\prime}}}w^{0},\qquad\tilde{w}^{1}=\frac{\pi_{1^{\prime}}}{\pi_{0^{\prime}}}w^{1}+\epsilon\left(w^{0}\right)^{3}\left(\frac{\pi_{1^{\prime}}}{\pi_{0^{\prime}}}\right)^{2}.
\ee
The holomorphic splitting of these relations, together with the Liouville 
theorem imply the existence of coordinates $(w,z,x,y)$  on $M_\C$  such that
the twistor curves pulled back from  ${\mathcal U}$ to ${\mathcal F}$ are
\[
w^{0}=w+y\lambda\qquad w^{1}=z-x\lambda-\epsilon y^{3}\lambda^{2}.
\]
The ASD Ricci--flat metric resulting from the Nonlinear Graviton construction
is of the form (\ref{pp_wave_m}) for $\gamma=-3\epsilon y^{2}$. We shall find 
the  connection on $M_\C$ directly. We find
\[
F_{\, B}^{A}=\begin{pmatrix}\lambda^{-1} & 0\\
3\epsilon\left(y\lambda+w\right)^{2}\lambda^{-2} & \lambda^{-1}
\end{pmatrix}, \quad
F_{\,00}^{1}=6\epsilon\left(y\lambda+w\right)\lambda^{-2},
\]
and all other $F_{\, BC}^{A}=0$.
The splitting problem (\ref{eq:SplittingProblem}) can 
be solved uniquely to give
\[
\tilde{\sigma}_{\,00}^{1}=-6\epsilon w, \quad\sigma_{\,00}^{1}=6\epsilon y, \quad 
\mbox{and all other}\,\,\{\sigma_{\, BC}^{A}\}=0,
\]
leading to a connection whose only non-vanishing components are
\[
\Gamma_{\, wy}^{x}=\Gamma_{\, yw}^{x}=-\Gamma_{\, ww}^{z}=-6\epsilon y.
\]
We have recovered  the Levi--Civita connection of (\ref{pp_wave_m}).
\subsection{Newtonian connections from $PT_{\infty}$}
We shall now construct the general class of connections
on the moduli space $M_\C$ of twistor lines in $PT_{\infty}=\OO\oplus\OO(2).$
Using (\ref{patch_U}) and (\ref{merpatch}) we find
\begin{equation}
F_{\, B}^{A}=\begin{pmatrix}1 & 0\\
0 & \lambda^{-2}
\end{pmatrix}\qquad\mbox{and}\qquad F_{\, BC}^{A}=0.
\end{equation}
The condition $F_{\, BC}^{A}=0$ implies that the connections will be constructed only
out of global sections of $N_{\mathcal{F}}\otimes\left(\odot^{2}N_{\mathcal{F}}^{*}\right)$.
In the relativistic case this would imply $\Gamma_{\, ab}^{c}=0$,
but for $PT_{\infty}$ we have $N_{\mathcal{F}}|_{L_p}=\mathcal{O}\oplus\mathcal{O}(2)$,
so
\begin{eqnarray*}
H^{0}\left(\mathbb{CP}^{1},(N_{\mathcal{F}}\otimes\left(\odot^{2}N_{\mathcal{F}}^{*}\right))_{L_p}\right)&=&
H^{0}\left(\mathbb{CP}^{1},\OO(-4)\right)\oplus H^{0}\left(\mathbb{CP}^{1},\OO(-2)\right)
\oplus H^{0}\left(\mathbb{CP}^{1},\OO(-2)\right)\\
&& \oplus \;H^{0}\left(\mathbb{CP}^{1},\OO\right) \oplus H^{0}\left(\mathbb{CP}^{1},\OO\right)
\oplus H^{0}\left(\mathbb{CP}^{1},\OO(2)\right)\\
&=&\C\oplus\C\oplus\C^3.
\end{eqnarray*}
Therefore we expect $\Gamma_{\, ab}^{c}$ to depend on five arbitrary
functions. 
The global sections of $N_{\mathcal{F}}\otimes\left(\odot^{2}N_{\mathcal{F}}^{*}\right)$
which constitute the Newtonian 0-cochain $\{\sigma, \tilde{\sigma} \}_{\infty}$
are of the form
\begin{gather}
\label{intermedia_eq}
\sigma_{\,TT}^{Q}=-\frac{1}{(\pi_{1'})^2}E^{A'B'}\pi_{A'}\pi_{B'}, \quad
\tilde{\sigma}_{\,TT}^{Q}=-\frac{1}{(\pi_{0'})^2}E^{A'B'}\pi_{A'}\pi_{B'}\nonumber \\
\tilde{\sigma}_{\, QT}^{Q}=\tilde{\sigma}_{\,T Q}^{Q}=\sigma_{\, QT}^{Q}=\sigma_{\,T Q}^{Q}=-i\chi \qquad\tilde{\sigma}_{\,TT}^{T}=\sigma_{\,TT}^{T}=-i\Sigma\,\,,
\end{gather}
(with all other components of $\{\sigma_{\, BC}^{A}\}_{\infty}$ set
to zero) for five arbitrary functions $(E^{A'B'}, \chi, \Sigma)$
on $M_\C$. Using (\ref{intermedia_eq}) and (\ref{ReadOffGamma}),
and imposing the reality conditions (\ref{newton_lines})
yields the
non-vanishing connection components
\begin{gather}
\Gamma_{\, \tau\tau}^{i}=E^{i}\qquad\Gamma_{\, \tau\tau}^{\tau}=\Sigma\qquad\Gamma_{\, j\tau}^{i}=
\Gamma_{\, \tau j}^{i}=\delta_{j}^{i}\chi,
\end{gather}
where we used the isomorphism (\ref{3spinor}) to 
replace $E^{A'B'}\in \Gamma(\spp'\odot\spp')$ by a spatial vector $E^i\in\Gamma(TM_\C)$.
The vector $E^{i}$ yields the gravitational  attraction, whilst $\Sigma$ and $\chi$ are consistent with 
$\nabla h=0$ and $\nabla\theta=0$, where the degenerate metric $h$, and
clock one--form $\theta$ are defined up to some conformal factors. 
When we fix these conformal factors as in Section \ref{sect_jump_lines},
so that $h=\mbox{diag}(0, 1, 1, 1,)$ and $\theta=d \tau$,  
then $\Sigma=\chi=0$. In particular, we note that this construction
cannot generate a Coriolis force.
What remains, then, is a connection whose non--zero $\Gamma_{\, \tau\tau}^{i}$
components are determined by a global section $\{\sigma_{\,TT}^{Q}\}_{\infty}$
of $\mathcal{O}(2)$. Using the Serre duality between the elements
of $H^1(\CP^1, \OO(-4))$ and $H^0(\CP^1, \OO(2))$ we can fix this connection by
\be
\label{serre1}
E_{A'B'}=\frac{1}{2\pi i} \oint_\Gamma g_{(-4)} \rho_{A'}\rho_{B'}\rho.d \rho,
\ee
where $g_{(-4)}$ is an element of $H^1(PT_\infty, \OO(-4))$ restricted to a twistor line
(\ref{rest_lines}). The resulting $E_{A'B'}$ satisfies the zero-rest-mass field equation
${\p^{A'}}_{C'}E_{A'B'}=0$. In what follows we shall demonstrate that the expression 
(\ref{serre1}) arises naturally from a limiting procedure applied to the unique splitting of the 0-cochain
corresponding to the Gibbons--Hawking metric. 
\subsection{Proof of Theorem \ref{theo_3}}
\label{sec_ward}
In Section  
\ref{GH_limit} we have demonstrated 
that
any Newtonian connection with the only non-zero components
given by $\Gamma_{\tau\tau}^i=\delta^{ij}\nabla_j V$ arises as a Newtonian limit
of the Gibbons--Hawking metric (\ref{gh_earlier}). The twistor space of
(\ref{gh_earlier}) is an affine line bundle over the total space
of $\OO(2)$ described by a cohomology class
$f\in H^1(\CP^1, \OO)$. This is in fact what we have recovered
in our analysis in Section \ref{instability} leading to the jumping phenomenon and the Kodaira instability of the Newtonian twistor space.
Let us assume that the cohomology class is represented by a function
$f=f(Q, \lambda, T)$ in a patch (\ref{patch_U}) in $PT_{c}$, and define
\be
\label{sec_ward1}
\widetilde{Q}=\frac{1}{\lambda^2}Q, \quad \widetilde{T}
=T-\frac{Q}{c\lambda}-\frac{1}{c^3}f,
\ee 
with $f=0$  giving the patching relation for the undeformed relativistic twistor space $PT_c$ (compare \ref{patch_flat}). We then have
\begin{eqnarray*}
F_{\, B}^{A}&=&\begin{pmatrix}1 & 
-\left(\frac{1}{c\lambda}+\frac{1}{c^{3}}\left[\frac{\partial f}{\partial Q}\right]_{\mathcal{F}}\right)\\
0 & \lambda^{-2}
\end{pmatrix}\,\,\,,\\
F_{\, AB}^{Q}&=&0\,\,\,\,\,\,\mbox{and}\,\,\, F_{\, AB}^{T}=\begin{pmatrix}F_{\,TT}^{T} & F_{\,T Q}^{T}\\
F_{\, QT}^{T} & F_{\, QQ}^{T}
\end{pmatrix}_{\mathcal F}=\begin{pmatrix}0 & 0\\
0 & -\frac{1}{c^{3}}\left[\frac{\partial^{2}f}{\partial Q^{2}}\right]_{\mathcal{F}}
\end{pmatrix}.
\end{eqnarray*}
For this twistor space we have $N_{\mathcal{F}}=\mathcal{O}(1)\oplus\mathcal{O}(1)$
(for finite, non--zero $c$) and so one can find the unique solution $\{\sigma, \tilde{\sigma}\}_{c}$
to the splitting problem (\ref{eq:SplittingProblem}). The solution is given by
\begin{eqnarray}
\label{IntegralQTT}
(\sigma_{\,TT}^{Q})_c&=&\frac{1}{2\pi i}\oint_{\Gamma}
\frac{1}{4(\pi_{1'})^2}
\frac{\p^2 f}{\p \omega^2}(\pi.\rho)^2 \rho. d\rho +\mathcal{O}\left(\frac{1}{c}\right)\\
(\tilde{\sigma}_{\,TT}^{Q})_c&=&
\Big(\frac{\pi_{1'}}{\pi_{0'}}\Big)^2 (\sigma_{\,TT}^{Q})_c\quad
\mbox{and all other}\quad\{\sigma, \tilde{\sigma} \}_c=\mathcal{O}\left(\frac{1}{c}\right),
\nonumber
\end{eqnarray}
where $\Gamma$ is a contour enclosing $\rho_{A'}=0$, and
$\omega=(1/2)(\rho_{1'})^2Q=\rho_{A'}\rho_{B'}x^{A'B'}$ is the global twistor function homogeneous of degree $2$ on $PT_\infty$ (compare formula (\ref{patch_U})) restricted to a twistor line. 
Thus one finds that
the only parts of $\{\sigma_{\, BC}^{A}\}_{c}$ which do not vanish
in the Newtonian limit are $\sigma_{\,TT}^{Q}$ and $\tilde{\sigma}_{\,TT}^{Q}$,
constituting a global section of $\mathcal{O}(2)$ and giving rise
to a non-zero $\Gamma_{\, \tau\tau}^{i}$ via Merkulov's procedure above.
This provides a way of fixing the Newtonian 0-cochain 
$\sigma_{\, BC}^{A}$:
we identify it with the $c\rightarrow\infty$ limit of the
Gibbons-Hawking 0-cochain
\be
\label{sigmaequalssigmainfty-1}
\{\sigma, \tilde{\sigma} \}_{\mathcal{\infty}}=
\lim_{c\rightarrow\infty} (\{\sigma, \tilde{\sigma}\}_{{c}}).
\ee
The only non--vanishing components are given by
$\sigma_{\,TT}^{Q}$ and
$\tilde{\sigma}_{\,TT}^{Q}$ in (\ref{intermedia_eq}),
where $E_{A'B'}$ is a zero-rest-mass field on $M_\C$.
Comparing  this limit with (\ref{serre1}) and replacing $\p_{A'B'}$ by $\rho_{A'}\rho_{B'}\p/\p \omega$
inside the integral yields
\begin{eqnarray*}
E_{A'B'}&=&-\frac{1}{2}\frac{\p}{\p x^{A'B'}} V, \quad\mbox{where}\\
V&=&\frac{1}{2\pi i} \oint_{\Gamma} \frac{1}{2}\frac{\p f}{\p \omega}\rho. d\rho
\end{eqnarray*}
where $f$ is an element of $H^1(PT_\infty, \OO)$ restricted to a twistor line
$\omega=x^{A'B'}\rho_{A'}\rho_{B'}$ and $V$
is the harmonic function which (up to a constant multiple)  gives the Newtonian potential in agreement with (\ref{GHcon}).

We conclude that the limiting procedure yields a canonical element of 
$H^1(PT_\infty, \OO(-4))$  in (\ref{serre1}). It is given by 
$g_{(-4)}=-(1/4)\p^2 f/\p \omega^2$. To complete the proof we shall show that
$f\in H^1(PT_\infty, \OO)$ gives rise to a line bundle 
$\nu:E\rightarrow PT_\infty$ which is trivial on twistor curves.
Let 
\[
{\mathcal U}=\{(Q, T, \pi_{A'}), \pi_{1'}\neq 0\}\quad\mbox{and}\quad 
\widetilde{\mathcal U}=\{(\tilde{Q}, T, \pi_{A'}), \pi_{0'}\neq 0\}
\]
be a covering of $PT_{\infty}$ and let
\[
\chi:\nu^{-1}(\mathcal U)\rightarrow {\mathcal U}\times\C, \quad
\tilde{\chi}:\nu^{-1}(\widetilde{\mathcal U})\rightarrow \widetilde{\mathcal U}\times\C
\]
be a local trivialisation of $E$. The holomorphic patching function
$F\equiv \tilde{\chi}\circ\chi^{-1}$ is given  by $F=e^f$. The multiplicative splitting of $F$ reduces to the additive splitting
(\ref{splitting_h}) of $f$, and it  gives rise to the Abelian monopole (\ref{ward_1}), where $(A, V)$ satisfy (\ref{monopole}). The 
harmonic function $V$ is identified with the Newtonian potential from Section \ref{newto_ex}. This construction gives all Newtonian connections with no Coriolis force, as in Section \ref{GH_limit} we have shown that all such connections arise as limits of metrics from the Gibbons--Hawking class.
\koniec
\subsubsection*{Example}
Consider a one-parameter family of Gibbons--Hawking metrics with linear 
potential \cite{DH,WM}
\[
g=(1+\epsilon z)(dx^2+dy^2+dz^2)+\frac{1}{\epsilon(1+\epsilon z)}
(d\tau +\epsilon^{3/2}xdy)^2,
\]
where $\epsilon=c^{-2}$.
This family is ASD and Ricci flat for all $\epsilon>0$. In the limiting
case $\epsilon\rightarrow 0$ the metric blows up, the inverse metric
degenerates to $\delta^{ij}$, but the Levi--Civita connection has
the finite limit with the only non-zero components given by
$
\Gamma_{\tau\tau}^z=1/2.
$
The cohomology class defining the line bundle from Theorem \ref{theo_3} 
\[
f=-\frac{\omega^2}{(\pi.o)^2(\pi.\iota)^2}
\]
gives rise to the monopole (\ref{twistor_monopole})
with
\[
V=z, \quad A=\frac{1}{2}(xdy-ydx).
\]
\section{Newton--Cartan connections from holomorphic vector 
bundles}
\label{sparling_section}
Theorem \ref{theo_3} gives a satisfactory twistor construction
of a Newtonian connection with no Coriolis term.
For a fixed degenerate metric $h$, 
the general Newton--Cartan connections correspond to closed two--forms $F$, which
(given a canonical 3+1 splitting induced by the one--form $\theta$) have 
well defined electric and magnetic parts 
(compare formulae (\ref{g_connection}) and (\ref{cor_con})). 
Thus we can identify
the connection with a couple of gauge fields on $M$, and construct it
from a generalisation of Ward transform \cite{ward} adapted to Newtonian 
settings. The gauge fields needed to reconstruct the gravitational and 
Coriolis parts of the connection consist of a couple of abelian monopoles.
The one--forms in the monopoles will give rise to an electric and magnetic 
fields on $M$, and these fields will in turn give rise to a connection.

This is a legitimate approach, as electromagnetic forces are inertial in the Newton-Cartan theory:
the equations of motion of a charged particle under the influence of the electric field ${\bf E}$ and magnetic field $2{\bf B}$
\be
\label{ini_em}
\ddot{\bf x}={\bf E}+2 {\bf B}\wedge \dot{\bf x}
\ee
can be reinterpreted as a geodesic motion corresponding to a
connection
\[
\Gamma_{00}^i=-E^i, \quad \Gamma_{0j}^i={\epsilon^i}_{kj}B^k,
\]
where now ${\bf E}$ is the gravitational acceleration, and ${\bf B}$ is the Coriolis force - compare (\ref{cor_con}). The field
${\bf B}$ is a gradient of a harmonic function $W$, and (assuming that 
${\bf B}$ is time--independent) ${\bf E}=-\nabla V$ is also a gradient, but 
$V$ is not necessarily harmonic. However $W^2+V$ is harmonic.

Let $M_\C$ be a moduli space of holomorphic sections of
the Newtonian twistor space
$PT_{\infty}=\OO\oplus\OO(2)\rightarrow \CP^1$, and let 
\[
{\mathcal F}=\{(p, \xi)\in M_\C\times PT_{\infty}, \xi\in L_p\}\subset M_\C\times PT_{\infty},
\]
be the five dimensional correspondence space. Here the {\em twistor line} $L_p=\CP^1$ is a rational curve in $PT_{\infty}$ with normal bundle
$\OO\oplus \OO(2)$ corresponding to $p\in M_\C$. The correspondence space can be identified with the projective spin bundle $\PP(\spp')$. 
Let $\mu:\spp'\rightarrow {\mathcal F}$ denote the corresponding fibration. 
Thus a pre-image of any twistor line in ${\mathcal F}$ can be pulled back to $\spp'$, where $\mu^*(L_p)=\C^2-\{0\}$.

Consider a vector bundle $E\rightarrow PT_{\infty}$ such that 
\[
E|_{L_p}={\OO(m)}\oplus \OO(n),
\]
where $n>m\geq -1$. Therefore the restriction of $E$ to a twistor line is a non-trivial vector bundle, and the pull back to a pre-image of $L_p$ in 
${\mathcal F}$ of the
patching matrix of $E$ can not be split
in a way which leads to the Ward twistor correspondence \cite{ward}. However pulling back $E$ to the total space of $\spp'$, and restricting it to a pre-image
$\C^2-\{0\}$ of $L_p$ in $\spp'$ is a trivial bundle. The corresponding pulled-back $GL(2, \C)$ patching matrix $F=F(Q, \pi_{A'}, T)$, where
$Q$ is given by (\ref{rest_lines}), satisfies\footnote{The splitting matrices
$H$  and $\widetilde{H}$ can be derived from those in (\ref{BH_split})
using
\[
\left(
\begin{array}{cc}
(\pi_{0'}/\pi_{1'})^{-m} & 0 \\
0 & (\pi_{0'}/\pi_{1'})^{-n}
\end{array}
\right) =
\left(
\begin{array}{cc}
(\pi_{0'})^{-m} & 0 \\
0 & (\pi_{0'})^{-n}
\end{array}
\right) 
\left(
\begin{array}{cc}
(\pi_{1'})^{-m} & 0 \\
0 & (\pi_{1'})^{-n}
\end{array}
\right)^{-1}.
\]
}
\[
F=\tilde{H} H^{-1}
\]
where $H:U\times M\rightarrow GL(2, \C)$ and $\tilde{H}: \tilde{U}\times M\rightarrow GL(2, \C)$ are holomorphic in $U$ and $\tilde{U}$ respectively, where
now $U$ and $\tilde{U}$ are pre-images of the standard covering of $L_p$ in $\C^2-\{0\}$. Note that $H$ and $\tilde{H}$ are not homogeneous on
$\C^2-\{0\}$ and so do not descend down to ${\mathcal F}$. Let $\nabla_{A'}:=\pi^{B'}\p/\p x^{(A'B')}$ be the a rank-two twistor ($\alpha$-plane) distribution on $\spp'$, such that the three--dimensional space of leaves of the rank-three distribution $\{\nabla_{0'}, \nabla_{1'}, \pi^{B'}/\p \pi^{B'}\}$ in $\spp'$ is $PT_{\infty}$. Therefore
$\nabla_{A'} F=0$, and the matrix components of $H^{-1}\nabla_{A'} H$ are homogeneous functions on the fibres of $\spp'$, with the coefficients given by functions on $M$. 
Let us rewrite this condition as
\be
\label{sparling_1}
\nabla_{A'} \left (
\begin{array}{cc}
v_{-m} & r_{-n} \\
w_{-m} & s_{-n}
\end{array}
\right )
 = \left (
\begin{array}{cc}
v_{-m} & r_{-n} \\
w_{-m} & s_{-n}
\end{array}
\right )    \left (
\begin{array}{cc}
\phi_{A'} & \delta_{A'} \\
\kappa_{A'} & \psi_{A'}
\end{array}
\right ),     \quad \mbox{where}\quad H= \left (
\begin{array}{cc}
v_{-m} & r_{-n} \\
w_{-m} & s_{-n}
\end{array}
\right ).
\ee
Here $(v_{-m}, r_{-n})$ and $(w_{-m}, s_{-n})$ denote a pair of local sections of $E$, where $(v_{-m}, w_{-m})$ are homogeneous of degree $-m$ and $(r_{-n}, s_{-n})$ 
are homogeneous of 
degree $-n$ when regarded as functions on the fibres of $\spp'\rightarrow M$. The polynomials $(\phi_{A'}, \psi_{A'}, \kappa_{A'}, \delta_{A'})$ will give rise to 
potentials for {\em higher spin fields} on $M$, i.e. sections of various powers of $\spp'\rightarrow M$ satisfying some field equations \cite{sparling}. To construct these fields, and find the corresponding field equations rewrite (\ref{sparling_1}) as
\begin{eqnarray}
\label{sparling_2}
\nabla_{A'} v_{-m}&=&v_{-m}\phi_{A'}+r_{-n}\kappa_{A'},\nonumber\\
\nabla_{A'} w_{-m}&=&w_{-m}\phi_{A'}+s_{-n}\kappa_{A'},\nonumber\\
\nabla_{A'} r_{-n}&=&v_{-m}\delta_{A'}+r_{-n}\psi_{A'},\nonumber\\
\nabla_{A'} s_{-n}&=&w_{-m}\delta_{A'}+s_{-n}\psi_{A'}.
\end{eqnarray}
Therefore, as $\nabla_{A'}$ is homogeneous of degree one,  
we conclude that $\phi_{A'}$ and $\psi_{A'}$ are homogeneous of degree one, $\kappa_{A'}$ is homogeneous of degree $(n-m+1) $, and finally $\delta_{A'}=0$. Equivalently
\[
\phi_{A'}=\phi_{A'B'}\pi^{B'}, \quad \psi_{A'}=\psi_{A'B'}\pi^{B'}, \quad \kappa_{A'}=\kappa_{A'B'\dots C'}\pi^{B'}\dots \pi^{C'}
\]
where
\begin{eqnarray*}
\phi\in\Gamma(\spp'\otimes\spp')&=&\Gamma(\mbox{Sym}^2(\spp')\oplus \C), \quad 
\psi\in\Gamma(\spp'\otimes\spp')=\Gamma(\mbox{Sym}^2(\spp')\oplus \C),\\
\kappa\in\Gamma(\spp'\otimes\mbox{Sym}^{n-m+1}(\spp'))&=&
\Gamma(\mbox{Sym}^{n-m+2}(\spp')\oplus \mbox{Sym}^{n-m}(\spp')).
\end{eqnarray*}
The $SL(2, \C)$ irreducible components give rise to two spin-2 fields, two functions, one spin-$(n-m)$  field and one spin $(n-m+2)$ field.
\[
\psi_{A'B'}=A_{(A'B')}+\ve_{A'B'}U, \quad \phi_{A'B'}=B_{(A'B')}+\ve_{A'B'}W, \quad \kappa_{A'B'C'\dots D'}=
\gamma_{(A'B'C'\dots D')} +\ve_{A'(B'}\rho_{C'\dots D')}.
\]
To find the field equations satisfied by these potentials, we shall impose the integrability conditions on (\ref{sparling_2}) arising from $\nabla_{A'}\nabla^{A'}=0$.
Contracting each equation in (\ref{sparling_2}) with $\nabla^{A'}$ and using (\ref{sparling_2})  to eliminate the derivatives of $H$ gives
\[
\nabla^{A'}\phi_{A'}=0, \quad \nabla^{A'}\psi_{A'}=0, \quad \nabla^{A'}\kappa_{A'}-(\phi^{A'}-\psi^{A'})\kappa_{A'}=0.
\]
Decomposing this last set of equations into irreducible parts yields the 
final system (which is the non--relativistic version of the Sparling equations 
\cite{sparling})
\begin{eqnarray}
\label{s_equations}
{\partial^{A'}}_{(C'} A_{B')A'}&=&\partial_{C'B'}U\nonumber\\
{\partial^{A'}}_{(C'} B_{B')A'}&=&\partial_{C'B'}W\\
{\partial^{A'}}_{(E'}\gamma_{B'C'D')A'}&=&\partial_{(B'E'}\rho_{C'D')}+
{\chi^{A'}}_{(E'}\gamma_{B'C'D')A'}+h\gamma_{B'C'D'E'} 
+\chi_{(B'E'}\rho_{C'D')},\nonumber
\end{eqnarray}
where $\chi_{A'B'}:=B_{A'B'}-A_{A'B'}$ and $h:=W-U$.
The first two equations give a pair of Abelian monopoles
$(A, U)$ and $(B, W)$ satisfying
\[
dA=*dU, \quad dB=*dW.
\]
Thus $U$ and $W$ are two harmonic functions, which give rise to electric and magnetic fields in (\ref{cor_con}) by 
\[
{\bf E}=-\nabla(U-W^2), \quad {\bf B}=\nabla W.
\]
If the gauge group is taken to be $SL(2, \C)$, then 
$\psi_{A'B'}=-\phi_{A'B'}$.
\subsubsection{A relation to Penrose-Ward correspondences}
Let $PT_c$ be a twistor space corresponding to a flat space-time
with the speed of light $c$. Thus, if $c\neq \infty$, then
$PT_c$ is the total space of a rank-two vector bundle $\OO(1)\oplus \OO(1)$,
and $PT_{\infty}$ is the $\OO\oplus \OO(2)$
Newtonian twistor space.

As a special case of the construction presented in the previous section, consider a vector bundle ${\mathcal E}\rightarrow PT_c$, such that
\[
{\mathcal E}|_{L_p}=N(L_p),
\]
where $N(L_p)$ is the holomorphic normal bundle to $L_p\subset PT_c$. Let
$\kappa$ be the canonical bundle of $PT_c$, so that 
$\kappa|_{L_p}=\OO(-4)$: this true both for finite, and infinite $c$
(as discussed in Section \ref{tf_def}).Define a rank-two vector bundle
\[
{E}=\kappa^{1/4}\otimes {\mathcal E}.
\]
If $c\neq \infty$, then ${E}$ is trivial when restricted to twistor lines, and its Ward correspondence gives a solution to anti-self-dual
Yang-Mills field on the complexified Minkowski space. In the Newtonian case
\[
{E}|_{L_p}=\OO(-1)\oplus \OO(1)
\]
corresponds to $n=1, m=-1$ in (\ref{sparling_1}). The resulting potentials
are two abelian monopoles, one spin-4 field and one spin-2 field
subject to equations (\ref{s_equations}). One can now identify the field strengths
of these potentials with various components of the Newton-Cartan connection.
Note that it is consistent to set $\gamma_{A'B'C'D'}=0$.

\section*{Appendix 1. Contour integral for 2+1 Schr\"odinger 
equation}
\appendix
\setcounter{equation}{0}
\label{sec_schrodinger}
\def\theequation{\thesection{A}\arabic{equation}}
It is known (see e.g. \cite{duval3}) that the free Schr\"odinger  equation in 
$(D, 1)$ space--time 
dimensions arises as a null reduction of the wave equation in $(D+1, 1)$
dimensions. We shall use this observation together with the twistor contour integral formula (\ref{penrose_form}) to construct a contour integral formula
for a $(2+1)$--dimensional Schr\"odinger equation. 

 Consider the wave equation (\ref{wave_eq}) which we shall write
as $\Box \phi=0$. Let $u=ct+z, v=ct-z$ be null coordinates, so that
$\Box=\p_x^2+\p_y^2-4\p_u\p_v$. Then
\be
\label{dirac_ansatz}
\phi(x, y, u, v)=e^{-\frac{imv}{2}}\psi(x, y, u)
\ee
satisfies $\Box \phi=0$ iff
\be
\label{schrodinger}
i\frac{\p\psi}{\p u}=-\frac{1}{2m}\Big(\frac{\p^2 \psi}{\p x^2}+
\frac{\p^2 \psi}{\p y^2}\Big)
\ee
which is the free Schr\"odinger equations in $2+1$ dimensions.
Now consider the integral formula (\ref{penrose_form}),
where $f=f(\omega^0, \omega^1, \lambda)$ is a twistor cohomology class
restricted to a twistor line (\ref{4d_incidence}).
The ansatz (\ref{dirac_ansatz}) implies that
$\p f/\p \omega^1=-imf/2$, where $\omega^1=(v+\lambda(x-iy))$.
Solving this equation for $f$, and substituting back into
(\ref{penrose_form}) yields the contour integral formula
\be
\label{schro_cont}
\psi(x, y, u)=\frac{1}{2\pi i}\oint_{\Gamma\subset \CP^1} e^{-\frac{1}{2}mi(x-iy)\lambda}
g(x+iy+\lambda u, \lambda)d\lambda,
\ee
where $g$ is an element of $H^1({\mathcal Z}, \OO(-2))$ restricted to a
line $L_p\cong\CP^1$ in ${\mathcal Z}=\OO(1)$. The complex two fold ${\mathcal Z}$ is
the twistor space of the flat holomorphic projective structure on a two--dimensional space
${\mathcal M}$. If $(\zeta, u)$ (with $\zeta=x+iy$)
 are local coordinates of a point  
$p\in {\mathcal M}$, then the corresponding line $L_p\subset {\mathcal Z}$
is given by $\eta=\zeta+\lambda u$, where $\eta$ is a coordinate on
the fibres of  ${\mathcal Z}\rightarrow \CP^1$. The formula
(\ref{schro_cont}) is also valid if $m=0$, where the real and imaginary parts of 
$\psi(\zeta)$ are harmonic functions on $\R^2$ depending on a 
parameter $u$. In this case (\ref{schro_cont}) is the Radon transform of $g$. 

 Let us give an example, where $m\neq 0$. Let $g(\eta, \lambda)=\eta^{-1}$, and 
let $\Gamma$ be a circle centred at the origin of the $\lambda$--plane.
Then
\begin{eqnarray*}
\psi&=&\frac{1}{2\pi i}\oint_{\Gamma}\frac{e^{-\frac{1}{2}mi(x-iy)\lambda}}{x+iy+\lambda u}d\lambda\\
    &=& \frac{1}{u}e^{\frac{im(x^2+y^2)}{2u}},
\end{eqnarray*}
which is indeed a solution of the $(2+1)$ Schr\"odinger equation
(\ref{schrodinger}).

The contour integral formula (\ref{schro_cont}) is, on dimensional grounds, 
only tangentially related to the rest of this paper where the non--relativistic theories in $(3+1)$--dimensions are considered 
(hence its place in the Appendix). It may however be relevant
in Newton--Cartan theories in $(2+1)$ dimensions \cite{ab, son2}.
\section*{Appendix 2. Spin connection as gauge field }
\appendix
\label{sec_app}
In the original Nonlinear Graviton construction \cite{penrose}, 
and its modification involving non--zero cosmological constant \cite{ward_wells}
the spin connection on $\spp\rightarrow M_\C$ can be constructed directly from the 
Ward correspondence applied to a certain rank-two 
sub-bundle of the tangent bundle of the twistor space. This construction is mentioned in \cite{ward_wells}, but not implemented explicitly. Below we show how  carry the construction over in a way, which at the linearised level, agrees with the spin connection
arising for Plebanski 2nd heavenly equations \cite{plebanski}.

Let $\kappa\rightarrow PT$ be the holomorphic canonical line bundle 
of the twistor space which restricts 
to $\OO(-4)$ on twistor lines.
If $PT$ corresponds to an ASD Einstein metric  (with or without $\Lambda$), then
there exists a one-form ${\mathcal T}$ (which is given by $\pi_{A'}d\pi^{A'}$ in the vacuum case.)
This one-form defines a rank-two 
sub-bundle ${\mathcal E}\subset T(PT)$ consisting of vectors  annihilated by 
${\mathcal T}$.
The bundle ${\mathcal E}$ restricts to $N$ on each twistor line.
The bundle ${\mathcal E}\otimes \kappa^{1/4}$ is therefore trivial on twistor lines, and 
(by the standard Ward transform \cite{ward}) corresponds to an ASD gauge field:
a spin connection on $\spp$. This agrees with the gauge field arising from the Sparling equation in Section \ref{sparling_section}.

Both spin bundles have a twistorial construction as
\[
\spp_p=\Gamma(L_p, {\mathcal E}\otimes \mathcal{O}(-1)), \quad
{\spp'}_p=\Gamma(L_p, \mathcal{O}(1)),
\]
where $L_p=\CP^1$ is a twistor curve corresponding to $p\in M_\C$.
Assume that the ASD metric on $M_\C$ is vacuum, and
so the twistor space fibres over $\CP^1$.
Let $f\in H^{1}(\CP^1, \OO(2))$.
To construction the connection on $\spp$ consider the inifinitesimal deformation
\be
\label{ap_ham}
\tilde{\omega}^A=\omega^A+\epsilon \frac{\p f}{\p \omega_A}
\ee
and let $U\in\Gamma({\mathcal E})$, so that $U=U^{AA'}\pi_{A'}\p/\p \omega^A$. The relation
\[
\tilde{\beta}^A\frac{\p}{\p\tilde{\omega}^A}={\beta}^A\frac{\p}{\p{\omega}^A}
\]
gives
\[
\tilde{\beta}^B={F^{B}}_A\beta^A, \quad\mbox{where}\quad
{F^B}_A={\delta^B}_A+\epsilon \frac{\p^2 f}{\p\omega_B\p \omega^A}.
\]
To construct the splitting $F=\widetilde{H} H^{-1}$ consider
\[
\widetilde{H}={\bf 1}+\epsilon\widetilde{G},\quad {H}={\bf 1}+\epsilon {G}
\]
so that, to the first order in $\epsilon$, 
\[
{\widetilde{G}_{A}}^B-{G_A}^B=\frac{\p^2 f}{\p\omega_B\p \omega^A}.
\]
The freedom in splitting the RHS is measured by elements
of $H^0(\CP^1, {\mathcal O})$. Choosing a constant spinor 
$\iota_{A'}$ yields
\[
{G_{B}}^C=\frac{1}{2\pi i}\oint_{\Gamma}\frac{\iota.\pi}{(\iota.\rho)(\pi.\rho)}
\frac{\p^2 f}{\p\omega_C\p \omega^B}\rho. d\rho.
\]
The usual Liouville argument gives
\[
H^{-1}\nabla_A H=\widetilde{H}^{-1}\nabla_A \widetilde{H}=\pi^{A'}\Gamma_{AA'}
\]
for some matrix $\Gamma_{AA'}$ which does not depend on $\pi_{A'}$.
Equivalently
\[
{({H^{-1})}_C}^D\nabla_A {H_D}^B=\epsilon{\delta_{C}}^D\nabla_A {G_{D}}^B
=\pi^{A'}{\Gamma_{AA'C}}^B.
\]
Therefore the linearised Ward transform gives a connection
\[
{\Gamma_{AA'B}}^C=
\iota_{A'}\frac{\epsilon}{2\pi i}\oint\frac{1}{(\iota.\rho)}
\frac{\p^3 f}{\p\omega_C\p \omega^B\p \omega^A}\rho. d\rho.
\]
This is consistent, in the linearised 2nd Plebanski gauge, with the expression
for the ASD Weyl spinor as \cite{DM}
\[
{C_{ABC}}^D={\p_A}^{A'}{\Gamma_{BA'C}}^D=\frac{\epsilon}{2\pi i}\oint 
\frac{\p^4 f}{\p\omega^A\p\omega^B\p \omega^C\p \omega_D}\rho.d\rho
\]
(in Plebanski's gauge the quadratic term in the connection
on $\spp$ contracts to zero, and the connection on $\spp'$ vanishes.).

\end{document}